\documentclass[preprint,3p,authoryear]{elsarticle}

\journal{Planetary \& Space Science}

\usepackage{natbib}

\usepackage{amssymb}

\usepackage{color}
\usepackage{doi}
\usepackage{hyperref}
\hypersetup{colorlinks=true, urlcolor=blue}

\begin{document}

\begin{frontmatter}



\title{Bouncing on Titan: Motion of the Huygens Probe in the Seconds After Landing\tnoteref{label1}\tnoteref{label2}}
\tnotetext[label1]{\doi{10.1016/j.pss.2012.08.007}}
\tnotetext[label2]{\copyright 2017. This manuscript version is made available under the CC-BY-NC-ND 4.0 licence:\\ \url{https://creativecommons.org/licenses/by-nc-nd/4.0/}}


\author[MPS]{Stefan E. Schr\"oder}
\author[LPL]{Erich Karkoschka}
\author[APL]{Ralph D. Lorenz}

\address[MPS]{Max-Planck-Institut f\"ur Sonnensystemforschung, 37191 Katlenburg-Lindau, Germany}
\address[LPL]{Lunar and Planetary Laboratory, University of Arizona, Tucson, AZ 85721-0092, U.S.A.}
\address[APL]{Applied Physics Laboratory, Johns Hopkins University, Laurel, MD 20723-6099, U.S.A.}

\begin{abstract}

While landing on Titan, several instruments onboard Huygens acquired measurements that indicate the probe did not immediately come to rest. Detailed knowledge of the probe's motion can provide insight into the nature of Titan's surface. Combining accelerometer data from the Huygens Atmospheric Structure Instrument (HASI) and the Surface Science Package (SSP) with photometry data from the Descent Imager/Spectral Radiometer (DISR) we develop a quantitative model to describe motion of the probe, and its interaction with the surface. The most likely scenario is the following. Upon impact, Huygens created a 12~cm deep hole in the surface of Titan. It bounced back, out of the hole onto the flat surface, after which it commenced a 30-40~cm long slide in the southward direction. The slide ended with the probe out of balance, tilted in the direction of DISR by around $10^\circ$. The probe then wobbled back and forth five times in the north-south direction, during which it probably encountered a 1-2~cm sized pebble. The SSP provides evidence for movement up to 10~s after impact. This scenario puts the following constraints on the physical properties of the surface. For the slide over the surface we determine a friction coefficient of 0.4. While this value is not necessarily representative for the surface itself due to the presence of protruding structures on the bottom of the probe, the dynamics appear to be consistent with a surface consistency of damp sand. Additionally, we find that spectral changes observed in the first four seconds after landing are consistent with a transient dust cloud, created by the impact of the turbulent wake behind the probe on the surface. The optical properties of the dust particles are consistent with those of Titan aerosols from Tomasko et al.\ ({\em P\&SS} {\bf 56}, 669). We suggest that the surface at the landing site was covered by a dust layer, possibly the 7~mm layer of fluffy material identified by Atkinson et al.\ ({\em Icarus} {\bf 210}, 843). The presence of a dust layer contrasts with the dampness measured just centimeters below the surface, and suggests a recent spell of dry weather at the landing site.

\end{abstract}


\end{frontmatter}



\section{Introduction}
\label{sec:introduction}

The Huygens probe entered Titan's atmosphere on January 14, 2005 and landed on its surface after a 2.5 hour descent\footnote{Strictly speaking, Huygens impacted onto the surface. Post-impact survival of the probe was not a design driver, and the mission would have been considered a success even if no further contact had been made. Fortunately, the impact turned out to be a soft landing, so we use the words ``impact'' and ``landing'' interchangeably.}. Images obtained by the Descent Imager/Spectral Radiometer (DISR) revealed the landing site to be a flood plain strewn with rounded, decimeter-sized pebbles of unknown composition \citep{T05,K07}. The surface itself appears fine-grained, damp, and relatively soft \citep{Z05,N05,L06b,K07}, and may be covered by a 7~mm thick fluffy layer \citep{A10}. The motion of Huygens during atmospheric entry and descent has been studied well \citep{B05,K07,L07}, but its motion during and directly following landing has not. In the last phase of the descent a stabilizer parachute restricted the vertical speed to 4.5~m~s$^{-1}$, ensuring a relatively soft landing. The horizontal speed was then only around 1~m~s$^{-1}$. It is clear that, upon impact, the 200~kg probe could not immediately have come to rest. Indeed, evidence for continued motion exists. \citet{S07} noticed that DISR measurements were variable for several seconds after impact, and suspected that motion and/or dust between the detector and the surface had been responsible. \citet{B08} analyzed Huygens Atmospheric Structure Instrument (HASI) accelerometer data, and proposed that the probe bounced back from the surface for 0.4~s with a horizontal speed of about 1 m~s$^{-1}$. The data indicated motion for a few more seconds after the bounce phase, which was not modeled. On the long term Huygens seems to have been very stable. In the remaining 70 minutes that Cassini was visible over the horizon it may have tilted very slowly, at a rate on the order of $0.1^\circ$ per hour \citep{Z05,K07}.

The goal of this paper is to develop a scenario for the events right after landing and to find a quantitative description for Huygens' motion. We hope our model not only fits the available data, but also offers insight into the physical properties of the surface. While the main focus is on the analysis of DISR data, we can only obtain a complete picture when we include data from other instruments. A variety of instruments onboard Huygens were active around impact. These can be roughly divided into two groups: accelerometers and photo/spectrometers. Accelerometers were part of HASI, the Surface Science Package (SSP), and the Radial Accelerometer Sensor Unit (RASU), whereas several photo- and spectrometers were included in DISR. The sparse data may not able to constrain all model parameters, and the solution may not be unique. Nevertheless, a quantitative model may allow a better understanding of Titan's surface near the landing site and of the dynamical interaction between the probe and surface. In addition, we look back at a drop test performed in Sweden with a Huygens model before launch, and assess whether it offers any clues to what happened on Titan.

This paper is divided into two sections. The first describes our analysis of the various data sets. The second describes the model we developed for the probe motion in the seconds after landing. We end our paper with a summary of our findings and discuss their implications for the physical properties of Titan's surface at the landing site. We have created several movies based on our motion model to visualize the events around landing. These, and a data file with the full model results are provided as online supplementary material.

\section{Data Analysis}

\begin{figure}
\centering
\includegraphics[width=10cm,angle=0]{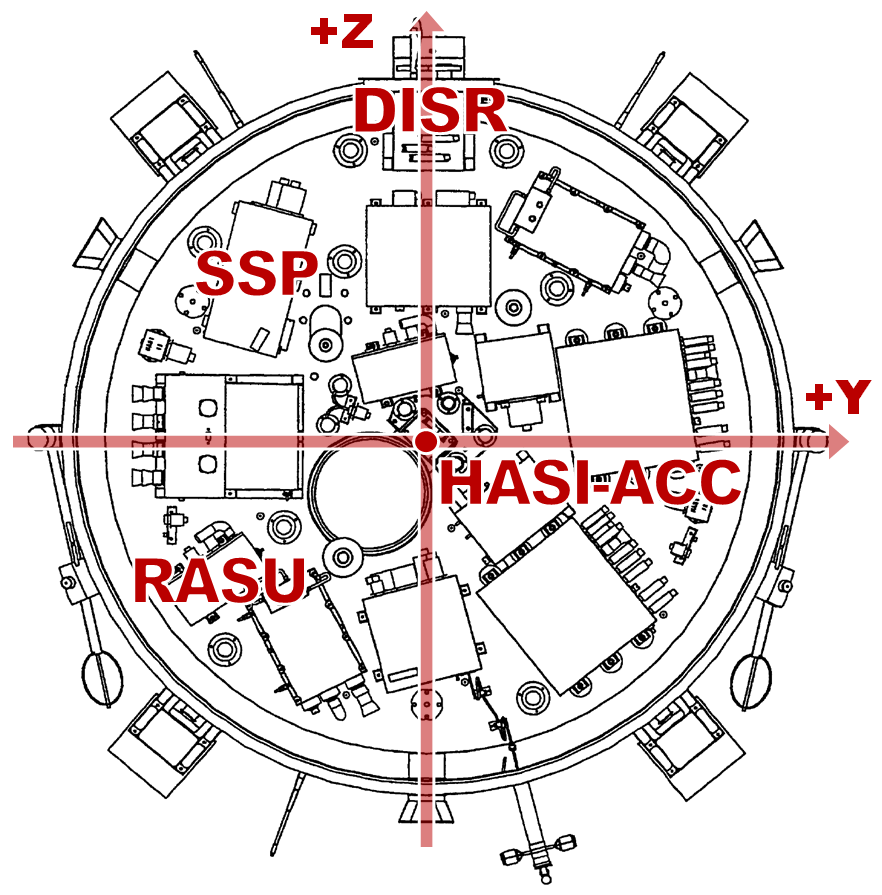}
\caption{Instrument locations on the Huygens instrument platform and definition of the probe $XYZ$-coordinate system. The $+X$-vector (not shown) points upward, towards the parachute. The location of the HASI accelerometers is indicated with a dot at the center of the platform. Probe diameter is 1.3~m. Image credit: ESA.}
\label{fig:Huygens}
\end{figure}

The data described in this section were mostly acquired by instruments located on the Huygens instrument platform inside the probe. We frequently refer to instruments by their acronym, listed in Table~\ref{tab:acronyms}. Their location on the platform is shown in Fig.~\ref{fig:Huygens}. This figure also defines the Huygens-centered coordinate system, in terms of which many of the data are discussed. The three axes are the $X$, $Y$, and $Z$-axis. The positive $X$-axis points up in the direction of the parachute, positive $Z$ points toward DISR, and positive $Y$ points $90^\circ$ to the right of DISR. We use ``mission time'', where impact is at mission time 8869.8~s. For exposures, the times quoted are mid-exposure.

\subsection{Descent Imager/Spectral Radiometer (DISR)}
\label{sec:DISR_analysis}

The Descent Imager / Spectral Radiometer (DISR) was mounted on the Huygens instrument platform aligned with the $Z$-axis (Fig.~\ref{fig:Huygens}). It carried imagers, spectrometers, and photometers to measure photometric properties of the atmosphere and surface \citep{T02}. In addition, DISR had a Surface Science Lamp (SSL), which was switched on before landing and stayed on. Several instruments recorded the spot on the surface illuminated by the lamp, in particular the Downward-Looking Visible Spectrometer (DLVS) and the Downward-Looking Violet photometer (DLV). The Medium Resolution Imager (MRI) also probed the lamp spot, but images were not taken around landing. However, one CCD column (\#49) was read out that probed the edge of the MRI image close to the lamp spot. The intensity distribution of the lamp spot can be approximated by a Gaussian, with a FWHM of $3.5^\circ$ (in the zenith-nadir direction). It is centered on nadir angle $20.5^\circ$.


Exposures of the DLV, DLVS, and column 49 were taken simultaneously about once per second around landing until 50~s past impact, when DISR switched to a different operation mode. These three data sets were transmitted to the Cassini spacecraft through two channels (A and B), but Cassini recorded only in one channel (B) \citep{LW05}. Thus, the available data are somewhat randomly spaced, and typically only data of one or two of the three instruments exist for any specific time. An overview of the data acquired is shown in Fig.~\ref{fig:DISR}. Calibration of the DLV and DLVS was described by \citet{K12}, and calibration of the data for column 49 was described by \citet{S07}. Below, we review the most important aspects for each instrument.

\subsubsection{Downward-Looking Violet Photometer (DLV)}

The Downward-Looking Violet Photometer (DLV) observed the spot illuminated by the Surface Science Lamp (SSL) with a wide field of view, almost $180^\circ$ in azimuth. Details about the calibration are described in \citet{K12}. The DLV made 12 instantaneous photometric measurements during the first 50~s after landing that were received on Earth \citep{S07}. The first two of these display different signal levels from the remaining 10 data points, which do not show any excursions beyond the estimated noise level. The measured intensity relative to the average of the last 10 data points is displayed in Fig.~\ref{fig:DISR} (top panel). The measured intensity depends on the lamp intensity, the reflectivity of the surface, the distance $d$ between the illuminated surface spot and the detector, and the phase angle $\varphi$ (solar illumination is negligible compared to the lamp output):
\begin{equation}
I = r(\varphi) F_{\rm SSL}(d).
\end{equation}
Here, the measured intensity $I$ has units W~m$^{-2}$~$\mu$m$^{-1}$~sr$^{-1}$ and the lamp irradiance $F_{\rm SSL}$ has units W~m$^{-2}$~$\mu$m$^{-1}$. The surface reflectivity $r$ is the bidirectional reflectance. We can convert the observed intensity into distance to the surface $d$ by making several assumptions. First, we assume that the lamp intensity remained constant. We use the result from \citet{K12} that spatial variations of the surface reflectivity around the landing site are minor, and assume constant reflectivity. We use the phase law from the same publication stating that the surface reflectivity changes roughly as $\varphi^{-0.1}$ for the phase angles in question. This changes the expected inverse square law for the intensity as function of distance to an intensity proportional to $d^{-1.9}$. For the final resting position, \citet{K07} determined an altitude of the window of DISR's Side Looking Imager above the ground of $48\pm2$~cm, a pitch angle of $-3.1\pm0.5^\circ$, and a roll angle of $0.9^\circ\pm0.5^\circ$. This means that DISR was looking up by $3.2^\circ\pm0.5^\circ$ with respect to a horizontal attitude. This angle is consistent with the HASI servo acceleration of $1.34300\pm0.00034$~m~s$^{-2}$ \citep{H09} when combined with the Titan reference gravity of 1.345~m~s$^{-2}$ \citep{L02}. Even though no uncertainty was given for the reference gravity, radar topography data show that Titan's radius at the landing site is within 0.02\% of the assumed 2575~km \citep{Z09}, and thus the gravity should be within 0.04\% of the reference value. Given this uncertainty, the HASI tilt angle is $3.1^\circ\pm0.6^\circ$. Using the slight positional offset between the instruments, the distance between the DLV and the illuminated surface spot was $50\pm2$~cm, and the same applies for the other detectors mentioned below. We will assume this number to be 50~cm in this work since our present work is focused on motion with respect to the resting position. If that number was really 48 or 52~cm, all distances will change by almost exactly the same amount, which leaves relative distances essentially unchanged. With this assumption, the relative intensity measurements of the DLV can be directly converted into distance data as shown on the right scale in Fig.~\ref{fig:DISR}.

\subsubsection{Column 49}

Light for DISR's visible spectrometers was recorded on the same CCD as light for the imagers, but at levels some 100 times fainter. DISR had no shutters. Thus, when spectra were acquired the CCD got flooded with light from the imagers, and the part of the CCD where spectra were recorded received some scattered light. In order to account for this, data in column number 49 were recorded because this column is located between the parts of the CCD recording images and spectra. Fig.~\ref{fig:MRI} displays an image recorded by the Medium Resolution Imager (MRI) after landing. The bright spot on the surface at bottom right is overexposed due to the strong illumination from the lamp. To the right of the MRI image on the CCD we find column 49, with rows 1-200 counted from the top down (Fig.~\ref{fig:MRI}). Column 49 registers scattered light, and as such measures the brightness of the image near its right edge. Row 200 received almost 100 times as much light as row~1 (Fig.~\ref{fig:Col49}, top panel). This is due to the fact that the reflector of the lamp focused light very strongly onto a relatively small spot, leaving the illumination further away fainter by several orders of magnitude.

\begin{figure}
\centering
\includegraphics[width=14cm,angle=0]{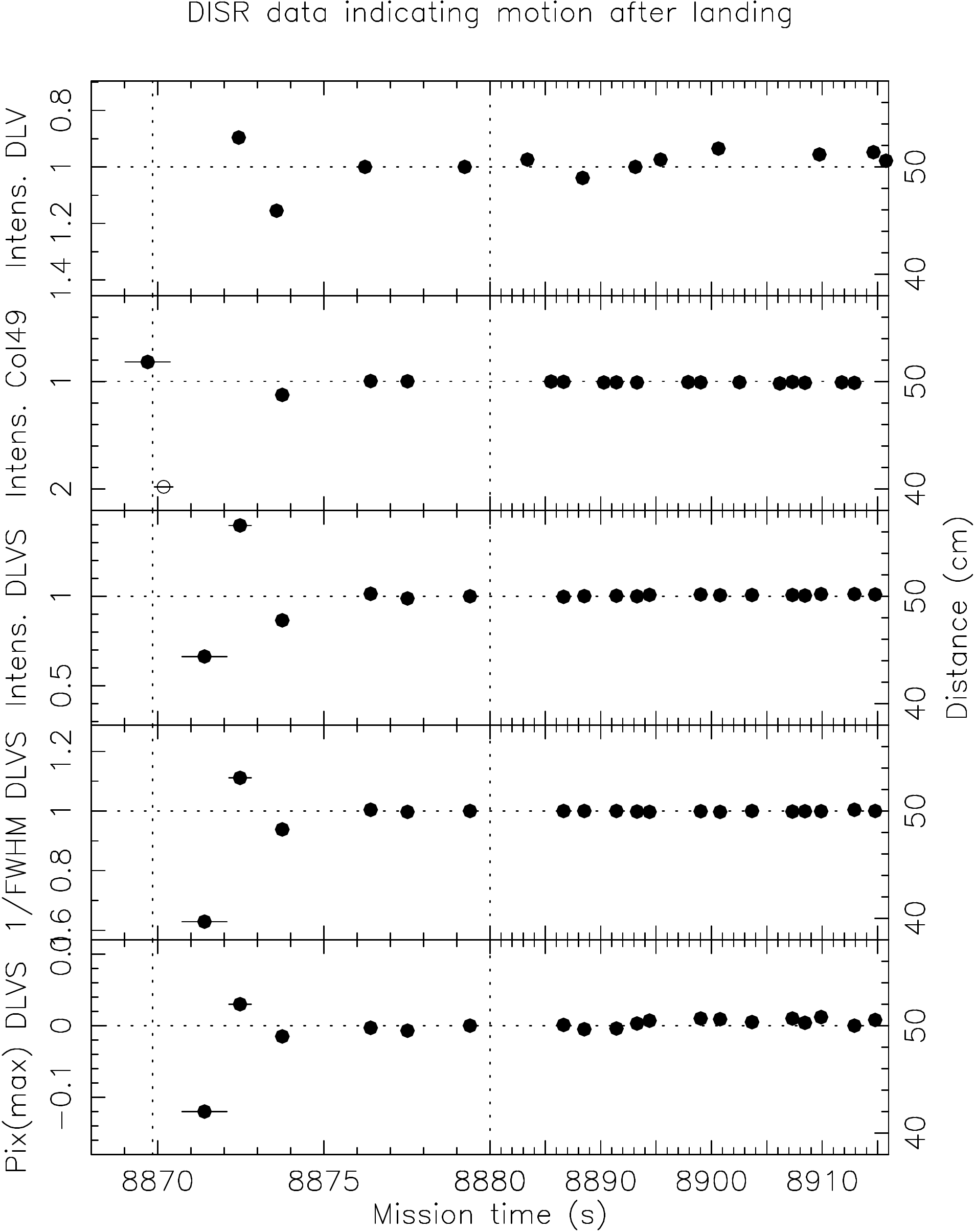}
\caption{The five measured parameters of DISR data during the seconds after impact relative to their values for Huygens at rest (scale at left). The intensity for column 49 was measured at row 190, and the intensity for the DLVS at 900~nm wavelength. Modeling of the data creates a relation between the data and the distance of the detector from the illuminated surface spot, which provides the scale at right. The time of landing is indicated with the vertical dotted line at the left of the figure. The open plot symbol for the column 49 intensity corresponds to our reconstruction for the post-landing signal only.}
\label{fig:DISR}
\end{figure}

During the first 50~s after landing, 17 exposures of column 49 were recorded. The exposures acquired around the time of landing exhibit intriguing variability \citep{S07}. The first of these was 1.36~s long and started already before landing, but ended after. Starting from the third exposure, all have essentially constant signal within the noise level. Thus, we averaged these 15 exposures and then normalized all data with respect to this average. We display smoothed, relative intensity curves in the bottom panel of Fig.~\ref{fig:Col49}. Even with the smoothing, the data are still noisy on the left side of the figure where original data numbers were low. Similar to the DLV, the intensity varies as $d^{-1.9}$, with $d$ the distance between the lamp spot and the detector. However, the changing parallax between the lamp and the detector causes an even larger variation, which changes the dependency to $d^{-4.3}$ close to the lamp spot, around row 190. As explained in \citet{K12}, our software can create model images for each requested distance. As the distance to the surface decreases, the parallax moves the bright spot further to the right in the image causing a large intensity increase at the high row numbers. We created model images for various distances and found a good fit to the second exposure (dashed curve in Fig.~\ref{fig:Col49}) for the probe rolled by $1^\circ$ toward the DISR side (i.e.\ a pitch-angle increase of $1^\circ$), corresponding to a distance decrease by 1.5~cm. The intensity at rows 1-120 increases by 5\% due to the $d^{-1.9}$ dependence. At higher row numbers, close to the lamp spot, the increase is larger due to the parallax shifting the lamp spot. Thus, we can convert the relative intensity measurements near row 190 (second panel in Fig.~\ref{fig:DISR}) into a distance scale, shown at the right side of Fig.~\ref{fig:DISR}.

The first (1.36~s long) exposure started at an altitude of 3.8~m when the illumination of column 49 was less than 2\% of the illumination after landing. Even at mid exposure, the illumination was still less than 20\%. The relatively small amount of illumination before landing can be estimated and subtracted from the data. This gives a measurement of the illumination for the part of the exposure that occurred after landing (open symbol in Fig.~\ref{fig:DISR}). Near row 190, the illumination was twice as bright as in the final resting position, toward row~1 almost three times as bright. During the first half second after landing, the detector must have been much closer to the surface than at the final position. We tried to model the shape of the solid curve in the bottom panel of Fig.~\ref{fig:Col49}, but we were unsuccessful.

While we believe that the detector was closer to the surface than in the final position, the quantitative connection of the intensity and distance as suggested in Fig.~\ref{fig:DISR} (left and right scale) is probably not accurate for this exposure. The intensity curve for the first exposure (solid curve in the top panel of Fig.~\ref{fig:Col49}) is much smoother than all others, which suggests major motion during the first half second after landing. The other 16 curves have all matching features that correlate with features seen in MRI images \citep{S07}. This implies that the horizontal motion of Huygens must have been quite small ($<$1~cm) from the time of the second exposure, 4.0~s after impact.

\subsubsection{Downward Looking Visual Spectrometer (DLVS)}

\begin{figure}
\centering
\includegraphics[width=8cm,angle=0]{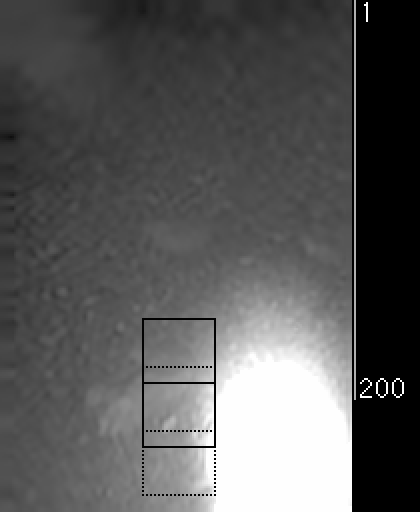}
\caption{A MRI image after landing. At bottom right is the overexposed spot on the surface illuminated by the lamp. The rectangles show the probed areas of the two spectra recorded by the DLVS, the solid rectangles for 600~nm, the dotted ones for 950~nm wavelength. The location of column 49 with rows 1-200 is indicated at right.}
\label{fig:MRI}
\end{figure}

\begin{figure}
\centering
\includegraphics[width=14cm,angle=0]{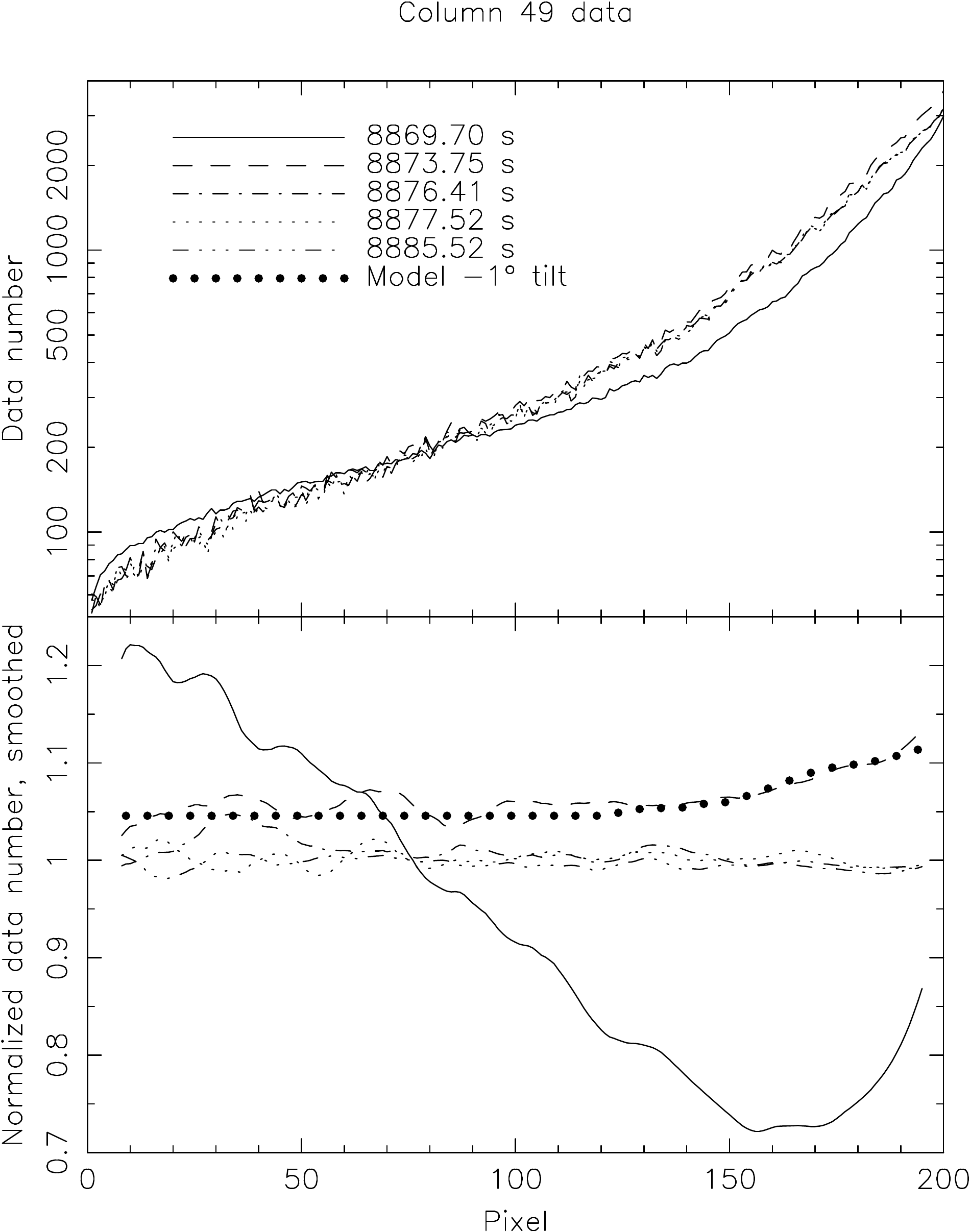}
\caption{The column 49 data for five exposures (top), and the smoothed data relative to the data for Huygens at rest (bottom). The dots show a calculation of the expected ratio curve for an $1^\circ$ increase in pitch angle of the probe, assuming a probe in contact with a flat surface. This calculation fits the data for the dashed curve at mission time 8873.75~s.}
\label{fig:Col49}
\end{figure}

The Downward Looking Visible Spectrometer (DLVS) recorded spectra containing 200 data points between 480 and 980~nm wavelength. During the first 50~s after landing, the DLVS acquired 19 exposures \citep{S07}. Each of these yields two spectra of the surface left of the lamp spot, one $4^\circ$ above the other (Fig.~\ref{fig:MRI}). In reality, the spectrum of a specific location along the slit was not perfectly aligned along columns on the CCD, but was angled and curved. Thus, for 600~nm wavelength, for example, the two probed locations correspond to the solid black rectangles in Fig.~\ref{fig:MRI}, while for 950~nm wavelength they correspond to the dotted rectangles. Using exposures acquired in the laboratory and on Titan, \citet{K12} showed that the 400 data points for each pair of spectra could be converted into 202 parameters, 200 of which correspond to a spectrum for the peak intensity along the slit. It was assumed that the shape of the spectrum does not change along the slit, based on spectra with 10 spatial resolution points along the slit taken minutes after landing. From these exposures it was also concluded that the spatial variation near the lamp spot can be approximated well by a Gaussian shape. The two remaining parameters then represent the spatial variation along the slit: the nadir angle of the peak intensity, and the full width at half maximum (FWHM) of the Gaussian. The unused 198 data points (400 minus 202) were used to check that the method and assumption were valid.

\begin{figure}
\centering
\includegraphics[angle=-90,width=14cm]{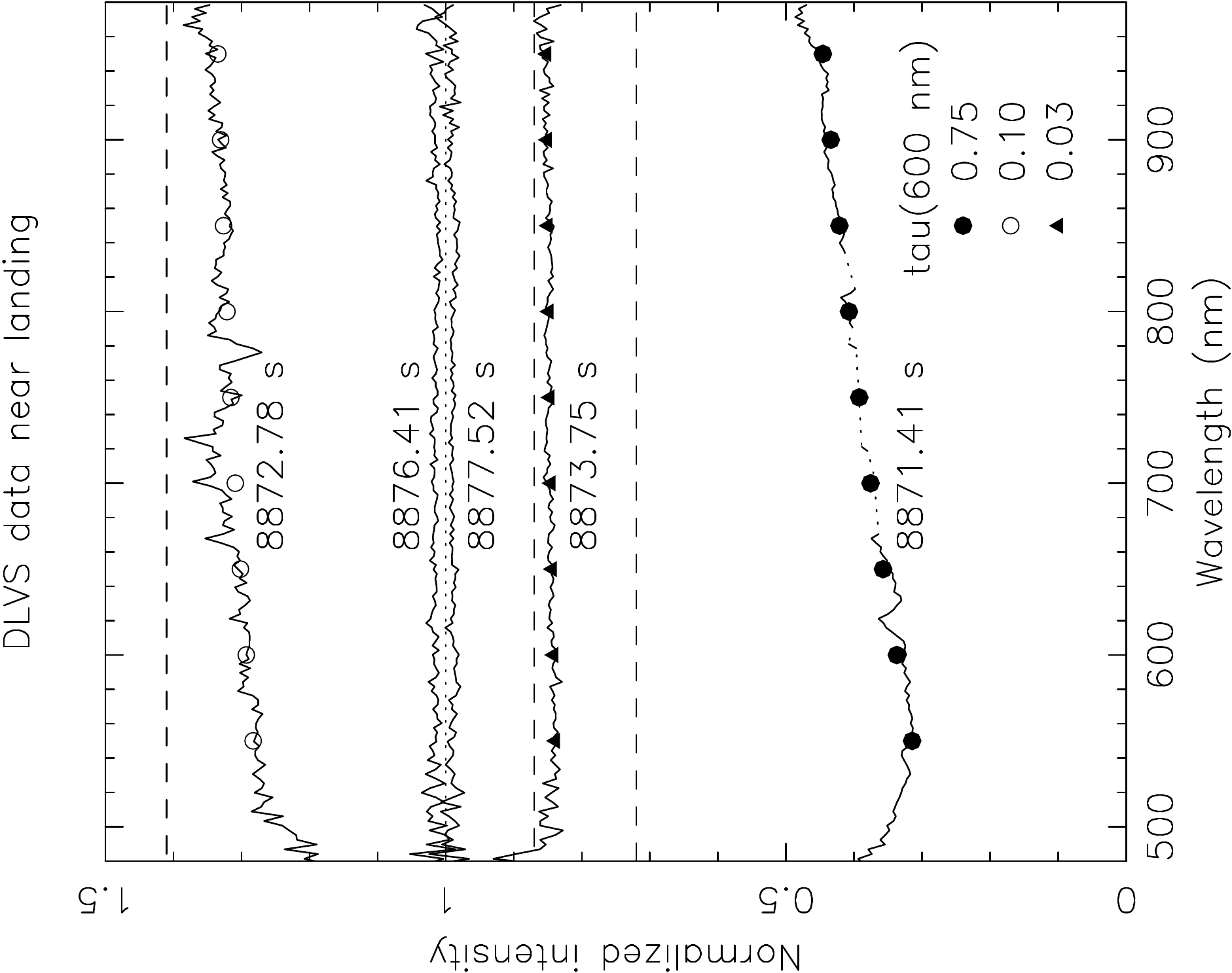}
\caption{The DLVS spectra for the first five exposures after impact, relative to the average spectrum for Huygens at rest. The sloping spectra can be explained by dust with absorption optical depth ($\tau$) inversely proportional to wavelength and $\tau$ at 600~nm decreasing with time as indicated. The dotted line is the normalization taken from later spectra. The dashed lines show the level for $\tau=0$ for the first three exposures.}
\label{fig:dust}
\end{figure}

The resulting spectra of the peak intensity along the slit are shown in Fig.~\ref{fig:dust} for the first five exposures. They are normalized to the average of the remaining 14 spectra since these were all identical within the noise level. Due to the relatively long exposure time, the first exposure (1.36~s long at 8871.41~s) is mostly saturated between 670 and 830~nm wavelength in both CCD columns (dotted line of bottom curve in Fig.~\ref{fig:dust}, linear interpolation). The exposure time of the next exposure (at 8872.78~s) was automatically decreased by the onboard software, and the central wavelengths are only saturated in the column probing the surface closest to the lamp spot. This spectrum was reconstructed using the fainter signal of the other column, but the result is obviously noisier at the central wavelengths (Fig.~\ref{fig:dust}, top curve). Subsequent spectra were all correctly exposed.

The basic intensity variation with changing distance is $d^{-1.9}$, as before. However, in this case the parallax works the opposite way as for column 49. As the distance decreases, the lamp spot moves to the right, away from the probed areas. This changes the intensity faster than the basic dependence. For distances around 50~cm, the intensity actually increases as $d^{2.2}$. Based on our modeling, the variable intensities measured during the first five exposures can be explained with variable distance to the surface, as shown in the central panel of Fig.~\ref{fig:DISR}. However, the model assumes that all spectra have the same shape, so the normalized spectra in Fig.~\ref{fig:dust} should all be flat and horizontal. This is not the case. The slope of the first three spectra taken after landing is steeper (redder) than that of later ones. In fact, the slope relaxes with a $1/e$ time constant of about 0.7~s. We consider several possible explanations.

Perhaps spatial variations across Titan's surface are responsible. Since our previous work suggests that the shapes of the visible spectra around the landing site are all almost identical \citep{K12}, we conclude that this is very unlikely. Then, the impact might have changed something in the instrument that came back to normal on the observed time scale. We consider this unlikely too.

Dust between the surface and the detector could have caused absorption with a temporal color variability. \citet{L93} speculated that sand or dust could be lofted for some time and might be detectable optically with DISR. \citet{A10} concluded that a 7~mm thick layer of a very fluffy substance lies on top of the surface at the landing site. It is natural to assume that this layer consists of settled aerosols. Huygens' impact might have suspended this layer in a dust cloud, which must have disappeared on a time scale shorter than a second. Since very small dust particles do not fall significantly within 0.7~s, they must have been carried away by the atmosphere. The wind speed at the surface was probably less than 0.25~m~s$^{-1}$ \citep{L06}, and possibly even lower at DISR's leeward side. If wind were responsible, the dust cloud would have been sized less than 15~cm to be consistent with the observed time scale. This is not credible. However, there is a turbulent wake behind a descending probe that dumps onto the ground as a ring vortex \citep{L93,E00}. The terminal velocity of sand-sized particles (0.2~mm diameter) at Titan is only 10~cm~s$^{-1}$ whereas the velocity of the probe at impact, and thus the characteristic velocity in the wake, is 5~m~s$^{-1}$. So, dust on the surface could form a small toroidal cloud with a length scale of about one probe radius and be advected radially outwards. The expected velocities are a little less than the descent speed (around 1~m~s$^{-1}$), consistent with the DLVS slope relaxation time. We model the spectra in Fig.~\ref{fig:dust} assuming a dust cloud with optical depth inversely proportional to wavelength, the same as found for aerosols below 30~km altitude \citep{T08}, and achieve an excellent fit. The derived optical depths given in the figure (0.75 at 600~nm for the first exposure) are for absorption along the double path from the lamp to the ground and back to DLVS.

Alternatively, the impact could have caused the lamp to temporally darken and decrease its color temperature. This hypothesis fits the data in Fig.~\ref{fig:dust} just as well as the dust model (not shown). The best fit equilibrium SSL color temperature is 2750~K. The color temperatures derived for the first three spectra after landing are 240, 40, and 10~K lower than equilibrium, which corresponds to a regular exponential decay with a factor of $1/e$ every 0.7~s. However, the first post-landing SSL voltage and current measurement (at 8874.09~s) does not show a significant change of power consumption. If a change in color temperature was caused by a very brief interruption of the power, then we cannot explain the slow return to the nominal color temperature. A typical light bulb like that of the SSL should respond much faster to switching on and off. We consider a change in color temperature an unlikely explanation for the DLVS slope change, and we favor the dust hypothesis.

What remains to be explained are the features at the left edge for the first two exposures (Fig.~\ref{fig:dust}). Actually, the upturn of the curve shortward of 550~nm wavelength for the first exposure and the downturn for the second exposure are expected. If the distance decreases, the parallax moves the lamp spot further to the right, which means that the observed area is illuminated by light rays further away from the lamp axis. These have less contribution from the gold-plated reflector of the lamp. Gold has almost constant reflectivity longward of 550~nm but lower reflectivity toward the blue part of the spectrum. Thus, the stronger contribution from direct light increases the blue illumination, as observed for the first exposure. The situation for the second exposure is just opposite.

The DLVS data provide us two other independent measures of the distance to the surface. We return to the two remaining parameters mentioned at the beginning of this section which describe the spatial variation of light along the slit: the width and nadir angle of the peak intensity. The width is estimated as the FWHM of a fitted Gaussian. Due to parallax, if the distance decreases, the lamp spot moves to the right in Fig.~\ref{fig:MRI}, and the sampled region far away from the spot has a broad, shallow maximum. For larger distances, the sampled area is close to the sharply peaked center of illumination. Between DLVS spectra taken before landing and those long after landing, the FWHM increases by 80\%, just as expected from our modeling. Thus, we are confident that we can convert the variation of FWHM recorded during the first exposures after landing into parallax variations and thus distance variations (fourth panel of Fig.~\ref{fig:DISR}, right scale). Because our model suggests that the inverse of the FWHM is almost linearly related to the distance between the detector and the ground, we plot the inverse of the FWHM on the left side of the fourth panel.

The nadir angle of the peak intensity can also be converted into a distance. The DLVS aperture is further away from the vertical axis of Huygens than the lamp. Thus, a changing distance has a small but measurable parallax effect in the vertical direction. We measured this effect between spectra taken before landing and those taken long after landing and found it consistent with our expectation from our model. Thus, we can convert the measured shift in the location of the peak intensity into distance changes (Fig.~\ref{fig:DISR}, bottom panel, right scale).

\subsubsection{Combined DISR data}
\label{sec:DISR_combi}

The data in Fig.~\ref{fig:DISR} suggest that around mission time 8871.5~s, DISR was closer to the surface by about 10~cm than in the resting position. One second later, all data suggest an increased distance with respect to the resting position by about 4~cm. Another second later, the sign changed again to a distance about 2~cm less than in the resting position. From this moment on, the data points of Fig.~\ref{fig:DISR} do not suggest any consistent motion. The noise associated with the DLV data is quite large, it is moderate for the DLVS peak location, and quite small for the remaining three parameters. The moderate disagreement between the distance data for the exposures before 8874~s is mostly due to uncertainties in the modeling. For some of the parameters where parallax changes the distribution of light, we do not expect our model predicts variations more accurately than 20\%.

\begin{figure}
\centering
\includegraphics[width=14cm,angle=0]{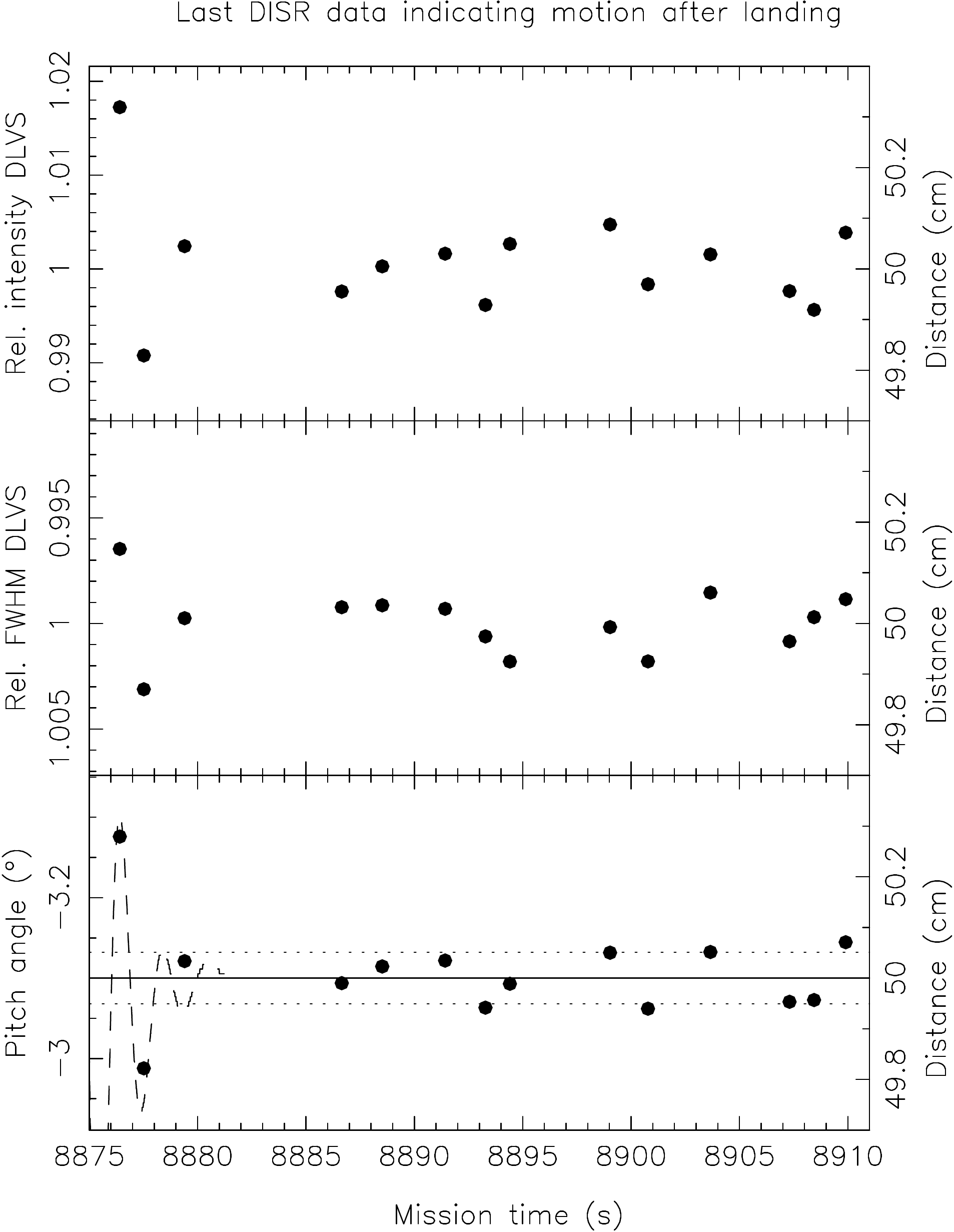}
\caption{A vertically expanded plot of the third and fourth panels of Fig.~\ref{fig:DISR} showing the DLVS intensity and FWHM measurements (top two panels) and the average of both data for the distance scale at right (bottom panel). The dotted lines signify the $1\sigma$ variation of data for Huygens at rest. The dashed curve is the model curve.}
\label{fig:DLVS}
\end{figure}

Considering that the noise is low for three parameters, in particular for the intensity and FWHM of the DLVS, we investigated whether we could detect any motion after 8874~s. In Fig.~\ref{fig:DLVS}, we show these data points in the top and middle panels. In order to decrease the noise, we averaged data numbers across the whole spectrum. The lower panel averages the distance estimates from both the top and middle panels. All data points after 8878~s are consistent with the estimated noise level of 0.05~cm, but the first two data points are off by $+0.3$ and $-0.2$~cm, respectively, which corresponds to $6\sigma$ and $4\sigma$. Thus, we conclude that Huygens still moved until 8878~s. Afterwards, the data are consistent with no motion. There is a slight chance that Huygens still moved when the data point at 8879.4~s was taken, but that this exposure was done when Huygens was half way between extremes. Thus, we cannot exclude motion during that exposure, but if Huygens was still in motion, the amplitude was probably very small.

Our data constrain only one parameter: the distance between the detector and the spot on Titan's surface illuminated by the lamp. The other five parameters describing the position and orientation of Huygens are unconstrained. Nevertheless, we can guess which type of motion occurred. During these observations, in particular in the later stages, Huygens was most likely constantly in contact with Titan's surface. Then, our measurements of the changing distance were due to the probe wobbling, changing the pitch angle in a damped oscillation. The roll and azimuth angle could have changed too, which would not be noticed in our data unless the variations were very large. Assuming that Huygens remained in contact with a flat surface, we convert the distance measurements into pitch angles (bottom left scale in Figs.~\ref{fig:DLVS} and \ref{fig:HASI}). The noise in our data corresponds to an uncertainty for variations in the pitch angle of $0.03^\circ$, or 2 arcminutes, which is a remarkably high precision.

The dust hypothesis is our preferred explanation for the slope changes observed for the DLVS spectra after impact. How would the presence of dust in the optical path length have affected the other DISR measurements? \citet{L93} calculated a dust cloud was possible based on the expected properties of the wake behind the probe. But the actual dynamics of the dust cloud are hard to predict. Dedicated computational fluid dynamics studies or scale model tests would be needed to identify the characteristic lengths and times of the turbulent dust cloud. By the time of first DLV exposure after landing, the dust was already quite subdued. The remaining dust should have absorbed some lamp light, but should also have scattered some lamp light in the backward direction. It is not clear which of these effects wins, so the DLV data cannot provide us any insight. For column 49 data, the exposure that started before landing may be affected by the presence of dust, but it is hard to know how. For the next exposure, the expected density of dust was already close to negligible. So, also the column 49 data are noncommittal. To minimize the effect of dust on the DLVS slope we use the data around 900~nm wavelength to quantify the DLVS intensity in Fig.~\ref{fig:DISR}. More dust increases the slope and decreases the intensity. This results in a lower inferred distance for the DLVS, a larger distance for the DLV, and no distance change for the other three measured parameters. Thus, averaged for all five parameters in Fig.~\ref{fig:DISR}, dust has a small effect on the distance scale, smaller than uncertainties from the modeling.

\subsection{Huygens Atmospheric Structure Instrument (HASI)}
\label{sec:HASI}

\begin{figure}
\centering
\includegraphics[width=14cm,angle=0]{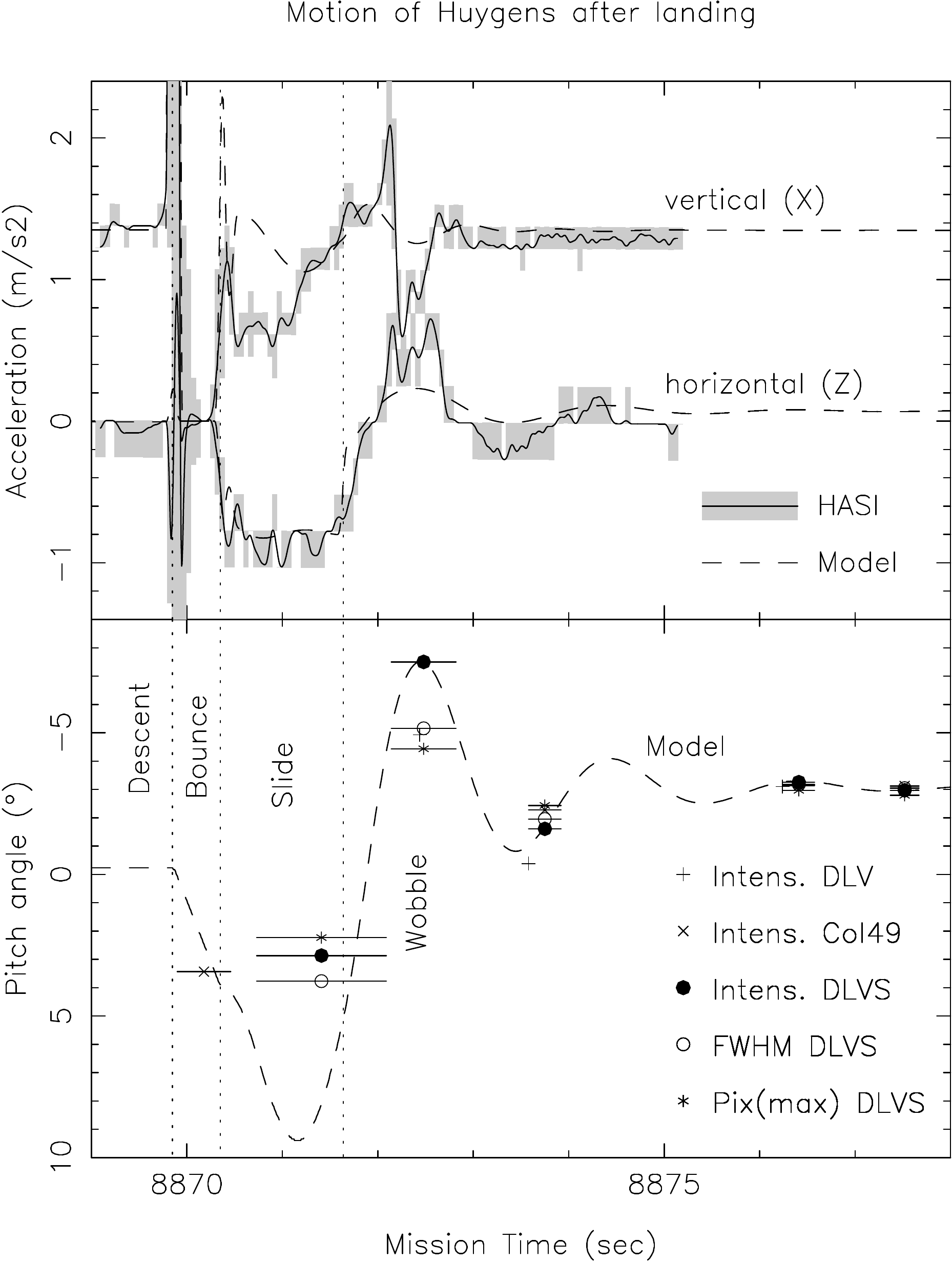}
\caption{Top: HASI accelerometer data for the $X$- and $Z$-direction (solid curves represent the data smoothed with a 0.06~s FWHM Gaussian, raw data in gray) compared with our model for Huygens' motion (dashed curves). Bottom: The pitch angle variation as inferred by DISR instruments (symbols) and the model curve (dashed).}
\label{fig:HASI}
\end{figure}

The Huygens Atmospheric Structure Instrument (HASI) piezo-accelerometers were mounted on the Huygens instrument platform near the probe center of mass \citep{F02} (Fig.~\ref{fig:Huygens}). They recorded the acceleration in all three directions of the Huygens $XYZ$-coordinate system, from 0.8~s before landing to 5.3~s after landing \citep{B08}. We retrieved the HASI data from NASA's Planetary Data System \citep{W08}. The data for the $X$- and $Z$-directions are plotted in Fig.~\ref{fig:HASI} (top panel) as solid curves\footnote{Note that there is uncertainty regarding the direction of the HASI $Y$-axis. Figure~1 in \citet{B08} shows the HASI axes aligned exactly with the Huygens $XYZ$-axes. On the other hand, the text reads {\em ``the sensing axes of the HASI accelerometers were all installed along the positive direction of the Huygens reference axes except for the piezo-$Y$ accelerometer, which faced the $Y$-direction''}, which meaning is not clear. We were unable to obtain clarification on this issue. We assume that the accelerations as shown in Fig.~2 in \citet{B08} are those in the Huygens $XYZ$- frame.}. Following \citet{B08} we plot positive $X$-acceleration values when the vector is pointed towards the negative $X$-direction (downward). In line with this convention, positive $Z$-accelerations point away from DISR. \citet{B08} postulated the existence of a period of free fall right after impact, resulting from a bounce. The raw data have a bias value, which we determined by forcing the acceleration to be zero during this bounce phase. Our bias determination could be off by 0.5 data number (DN) for each curve, and the digitization for all data points could also cause an error by 0.5~DN. Thus, our goal is to develop a motion model that can fit the data within 1~DN, which is indicated by the vertical width of the gray areas in Fig.~\ref{fig:HASI}. \citet{B08} integrated the HASI data and reconstructed a horizontal motion of about 1~m~s$^{-1}$ in the $Z$-direction during the bounce phase, followed by a sliding, decelerating motion on the surface. The primary impact at mission time 8869.8~s caused a maximum vertical acceleration of 120~m~s$^{-2}$ (off-scale in Fig.~\ref{fig:HASI}), due to the probe's deceleration within 12~cm \citep{Z05}. The authors considered two possibilities; that (1) Huygens either created a hole and bounced out of it, or that (2) it displaced a pebble on the surface and ceased vertical motion just at ground level. We consider the second possibility unlikely. Our new result from Sec~\ref{sec:DISR_analysis} is that Huygens could not have been perched on a pebble because DISR was very close to the surface during the first half second after impact. Also, data from the penetrometer of the Surface Science Package \citep{A10} might be difficult to explain. Furthermore, the secondary impact after the bounce phase would have occurred with an impact speed of 0.27~m~s$^{-1}$, while the HASI $X$-acceleration spike at that time suggests a vertical speed change of only 0.05~m~s$^{-1}$. All these observations can be better explained with the first possibility, which we adopted for our motion model in Sec.~\ref{sec:motion_model}. Thus, for the rest of this work, we assume that Huygens made first contact with Titan at the ground level, which implies that it was digging a hole of 12~cm depth.

\begin{figure}
\centering
\includegraphics[width=14cm,angle=0]{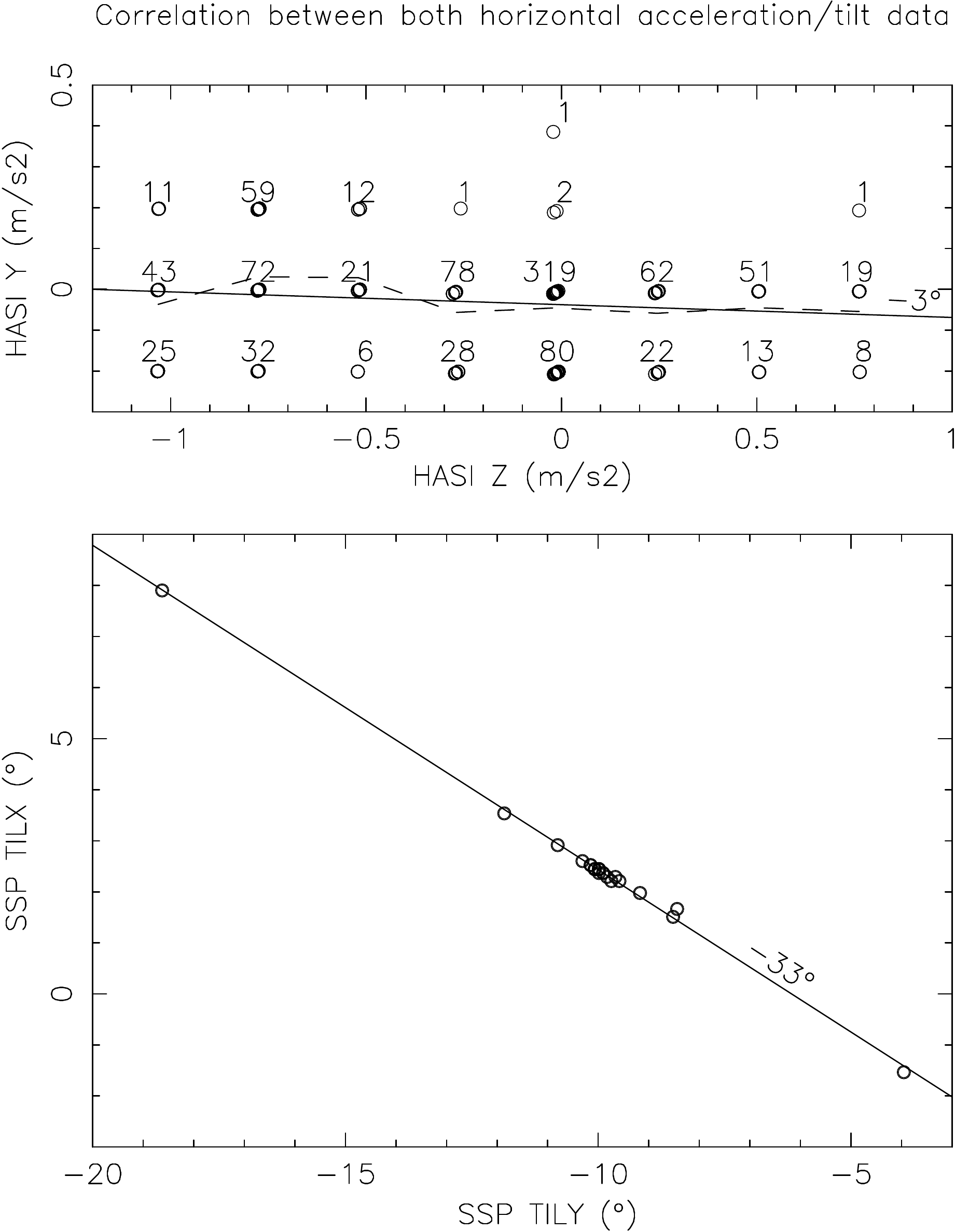}
\caption{HASI acceleration data (top) and SSP tilt data (bottom) for both horizontal directions during the slide and wobble phase. The 1046 HASI data points are grouped as 23 discrete data points. The data are quantized due to the relatively low resolution of the HASI piezo-accelerometers. The number above each data point displays the frequency of occurrence for that measurement. The dashed line connects averages for each of the eight vertically aligned groups. The solid line is a linear fit, corresponding to a direction of motion $3^\circ$ left of the $Z$-axis. The best linear fit for the tilt data has a slope of $-33^\circ$ degrees, which also corresponds to a motion vector $3^\circ$ degrees left of the $Z$-axis considering that the SSP instrument was rotated by $30^\circ$ degrees with respect to the principal axes of Huygens (Fig.~\ref{fig:Huygens}).}
\label{fig:motion_vector}
\end{figure}

After the primary impact, the probe had a zero-$g$ phase for a duration of 0.4~s, indicating a bounce with no surface contact \citep{B08}. Subsequent $X$- and $Z$-accelerations had some significant variations for a few seconds until settling down. The $Y$-acceleration on the other hand was 70\% near zero, 10\% at $+1$~DN, and 20\% at $-1$~DN (1~DN corresponds to about 0.2~m~s$^{-2}$). These low signals indicate that most of the motion was in the $X$-$Z$ plane (c.f.\ Sec.~\ref{sec:AGC}). By correlating the $Y$- and $Z$-data we determine an average motion vector of about $3^\circ$ to the left of the $Z$-axis (Fig.~\ref{fig:motion_vector}). Using the azimuth of $193\pm5^\circ$ ($13^\circ$ west of south) for the $Z$-axis of Huygens at rest from \citet{K07}, the motion vector of Huygens was around azimuth $190^\circ$, very close to pointing south, consistent with SSP measurements (see Sec.~\ref{sec:SSP}).

After the bounce and slide phase, Huygens wobbled on the surface for several seconds. This was already inferred from DISR data (Sec.~\ref{sec:DISR_combi}), but the HASI data allow us to model it in detail as a damped oscillation. This model, the particulars of which are described in Sec.~\ref{sec:motion_model}, is drawn in the top panel of Fig.~\ref{fig:HASI}. In the bottom panel, we see how the probe pitch angles derived from DISR data in Sec.~\ref{sec:DISR_combi} compare to the model predictions. Not all model parameters are derived from HASI data exclusively; the damping time constant of the oscillation is based on DISR observations because the HASI data do not extend beyond 8875~s. The DISR-derived pitch angles provide an independent test for our motion model, and the good fit of the data in the bottom panel in Fig.~\ref{fig:HASI} may convince the reader that it has passed the test.

\subsection{Surface Science Package (SSP)}
\label{sec:SSP}

\begin{figure}
\centering
\includegraphics[width=12cm,angle=0]{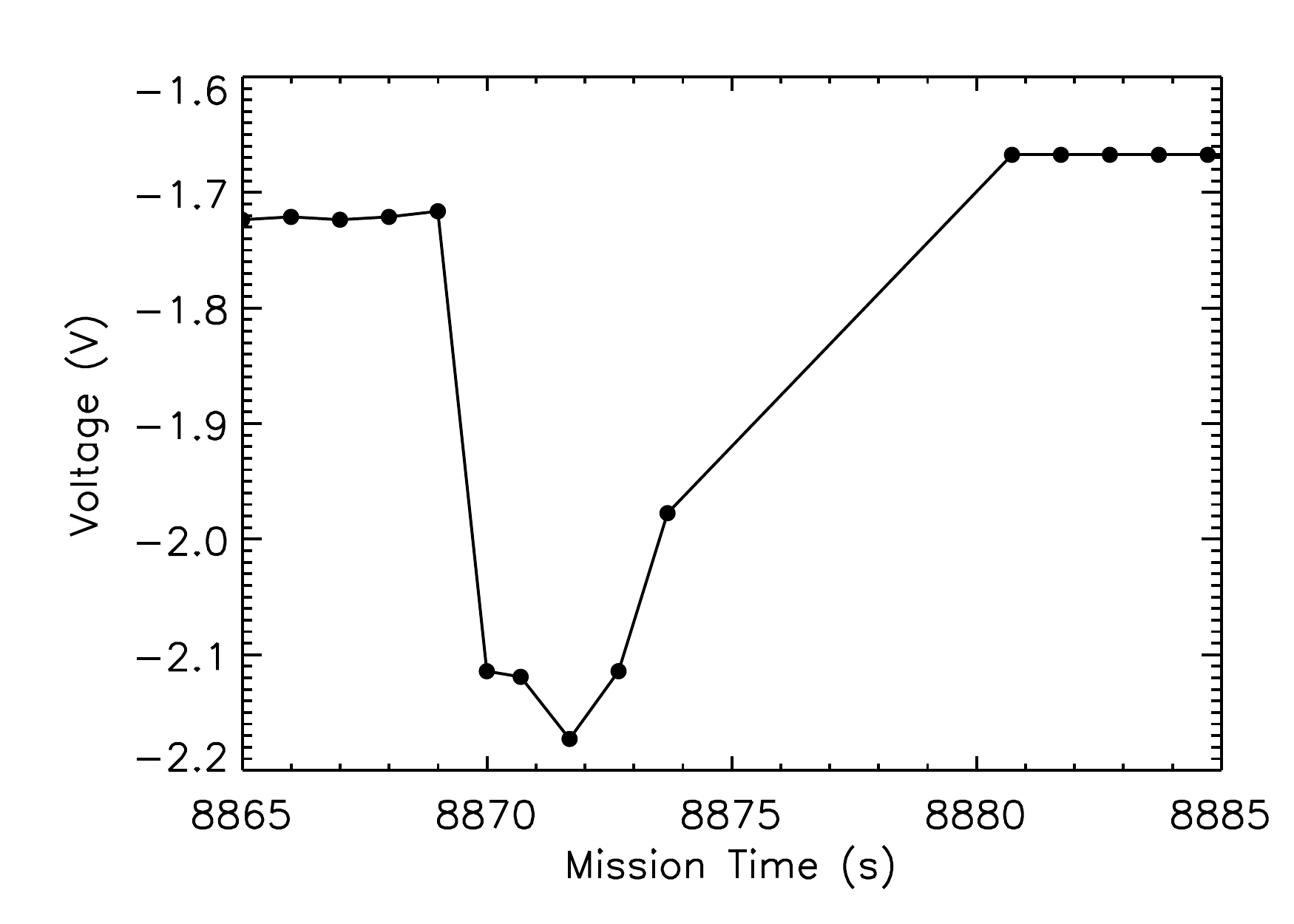}
\caption{Signal recorded by the SSP density sensor around the time of impact.}
\label{fig:SSP_DEN}
\end{figure}

The Surface Science Package (SSP) was mounted on the instrument platform rotated by $30^\circ$ with respect to the Huygens $Y$ and $Z$-axes \citep{Z02} (Fig.~\ref{fig:Huygens}). Continued motion after impact is also supported by data from SSP sensors \citep{Z05}. One of these is a density sensor, a small float cantilevered on a strain gauge, which fortuitously also functioned as a crude accelerometer \citep{L07}. Sampling of this sensor was interrupted shortly after landing while a packet of priority data was assembled in the instrument \citep{L12}, but the timed data (Fig.~\ref{fig:SSP_DEN}) shows motions or vibrations at least until mission time 8873.6~s (4~s after impact). The signal has returned to quiescent levels by 8880~s. This is suggestive of continued motion, even though the possibility that the sensor experienced relaxation cannot be excluded. The others are two radially oriented tilt sensors, used to monitor the descent \citep{L07}. These tilt sensors measured the position of a small slug of conductive fluid in a cylindrical vial, thus (in steady state) giving an indication of the orientation relative to the gravity field. In a dynamic environment they indicate lateral accelerations.

\begin{figure}
\centering
\includegraphics[width=12cm,angle=0]{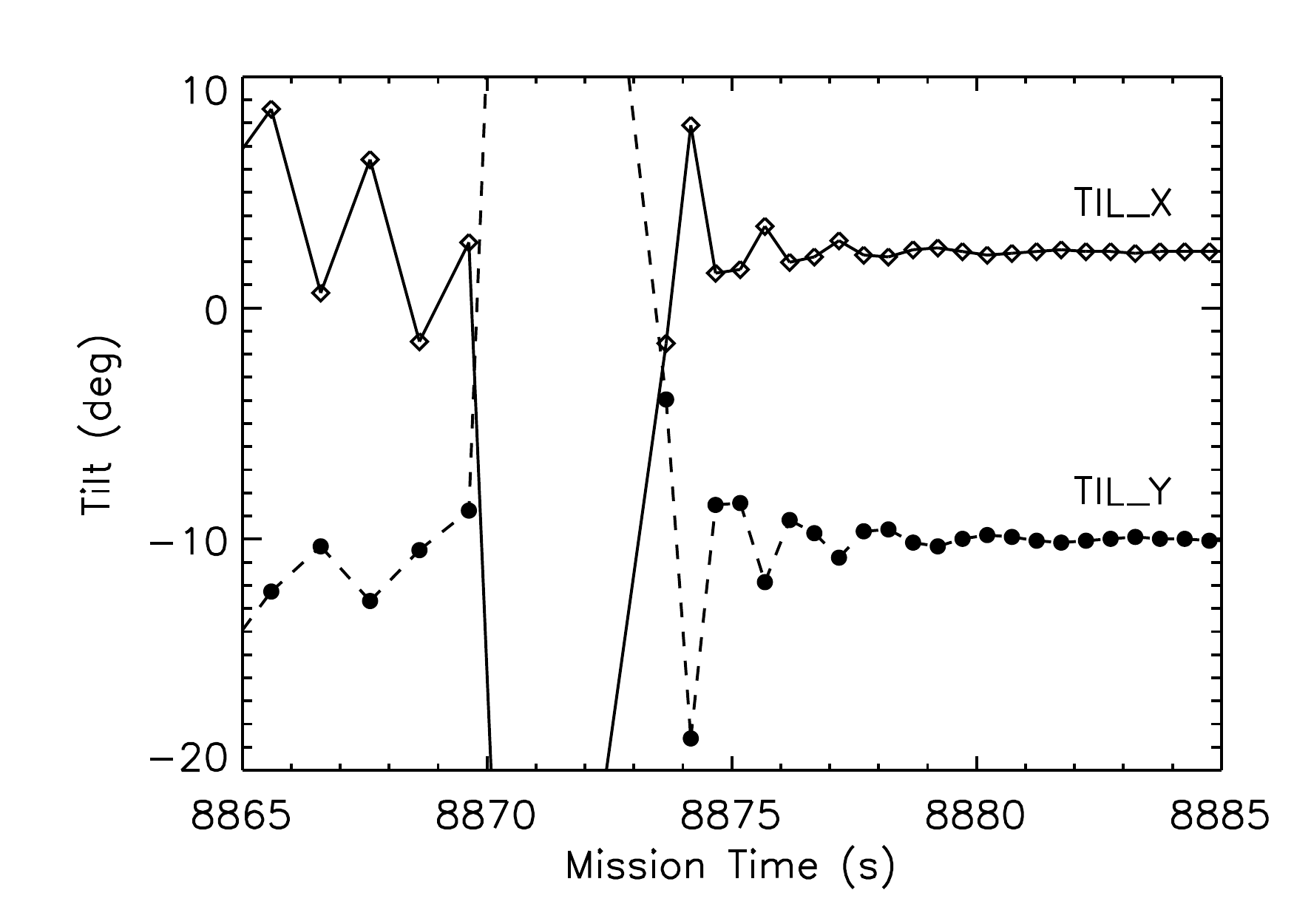}
\caption{Huygens probe tilt around the time of impact as measured by the two SSP tilt sensors TIL\_X and TIL\_Y.}
\label{fig:SSP_TIL}
\end{figure}

The time stamps on the SSP tilt records as archived on the PDS are in error (though may be corrected in due course). We used corrected timing of the tilt data by Leese and Hathi (private communication) to make it consistent with that of the DISR and HASI data sets. We restrict the tilt angle range in Fig.~\ref{fig:SSP_TIL} to clearly show the subtle tilt variations recorded a few seconds after impact, which appear to indicate that the probe performed a damped oscillation or wobble. A maximum occurs near mission time 8874.2~s, with a total tilt of about $10^\circ$ relative to the stable position. This tilt corresponds to a $Z$-acceleration of $+0.2$~m~s$^{-2}$ (Titan's gravity multiplied by $\sin 10^\circ$), which is consistent with the actual HASI reading within the quantization (Fig.~\ref{fig:HASI}). Our model predicts a pitch angle of $-1^\circ$ relative to the stable position. This implies that about 90\% of the tilt data during the wobble was due to the acceleration of the probe, and only 10\% due to the tilt of the probe. Soon afterwards, HASI stopped recording. DISR recorded another two cycles of the wobble, while SSP covered 3-4 more cycles. So, including the 1.5 cycles of wobble before 8874.2~s, we have evidence for five complete cycles of wobble. The second maximum of the fifth cycle, or the tenth maximum of the wobble, occurred near mission time 8880~s with a tilt of $0.2^\circ$ off the stable value, corresponding to a probe pitch angle $+0.02^\circ$ off the stable position.

The data from the two tilt sensors in Fig.~\ref{fig:SSP_TIL} are strongly correlated after impact, indicating that the wobble motion was planar. The bottom panel in Fig.~\ref{fig:motion_vector} shows that this correlation corresponds to an average motion in a direction $3^\circ$ left of the $Z$-axis, exactly as determined from HASI data (top panel; see Sec.~\ref{sec:HASI}). This remarkable agreement is a testament to the high quality of the data of both instruments.

\subsection{Radial Accelerometer Sensor Unit (RASU)}

\begin{figure}
\centering
\includegraphics[width=12cm,angle=0]{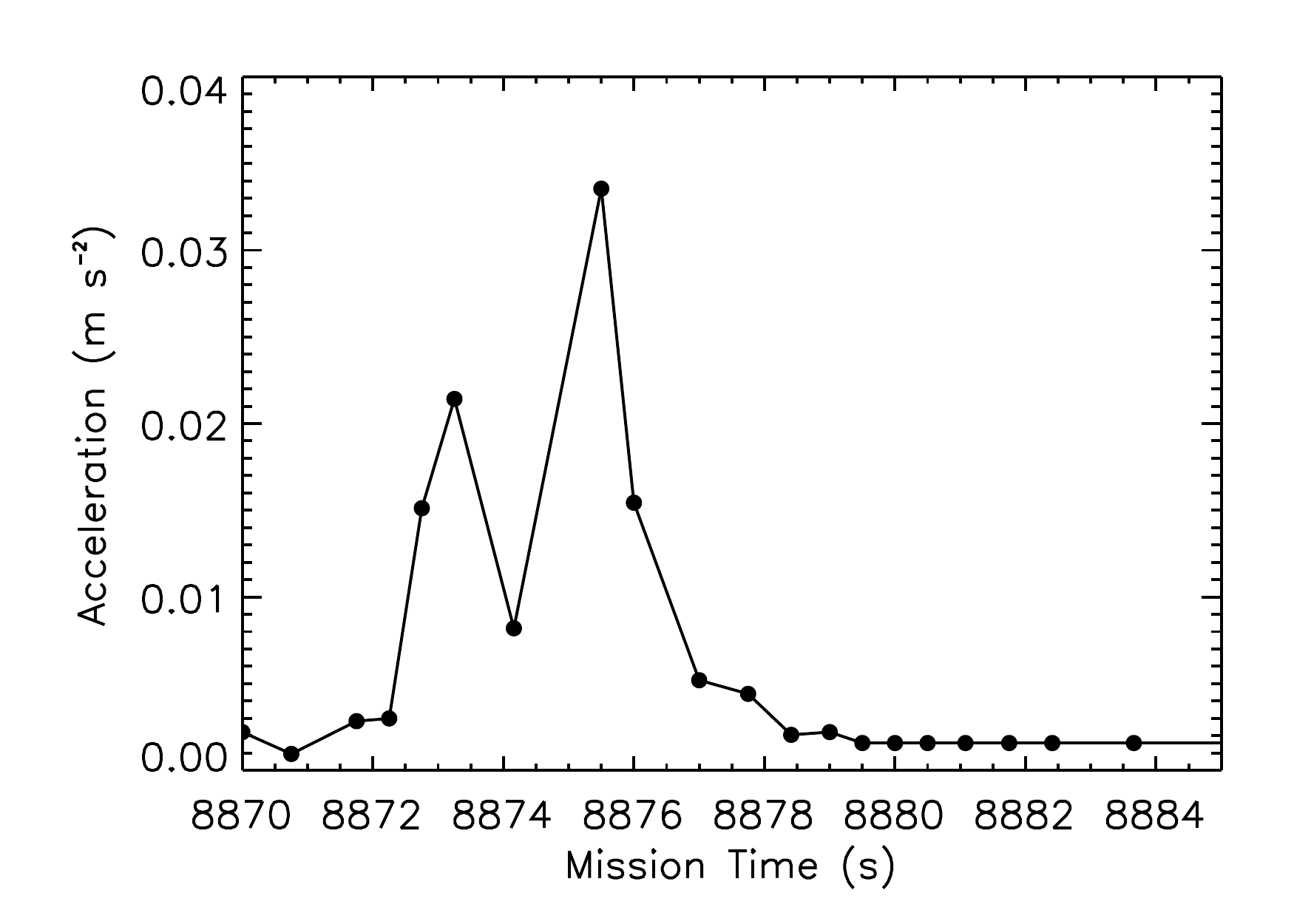}
\caption{Accelerations measured around the time of impact by RASU accelerometer~3.}
\label{fig:RASU}
\end{figure}

The Huygens probe carried the Radial Accelerometer Sensor Unit (RASU) to measure the probe spin rate (Fig.~\ref{fig:Huygens} shows its location on the instrument platform). RASU data was telemetered at irregular intervals. These data, which also shed light on turbulent motions of the probe during descent \citep{L07}, are archived on the PDS {\em Atmospheres Node} in the Huygens housekeeping data directory\footnote{Data set ID is HK\_CDMS\_RASU\_D8005A.TAB}. The data around impact from one of the accelerometers are shown in Fig.~\ref{fig:RASU}. The first signs of impact are only evident after 8872~s, which suggests that RASU time is offset by about 2~s from mission time. The signal does not return to quiescent levels until 8879.5~s, suggesting a combination of probe motion and/or structural ringing persisting for at least 7~s after impact.

\subsection{Automatic Gain Control (AGC)}
\label{sec:AGC}

The gain of the signal from Huygens received by Cassini, the Automatic Gain Control (AGC), was measured at a frequency of 8~Hz. Before landing it varied due to the spin of Huygens at a rate around 1~RPM \citep{K07}. After landing, it continued to vary for a few seconds due to probe motion. Since Cassini was located roughly in the direction of the $Y$-axis of Huygens, the main part of the motion in the $X$-$Z$ plane would not have caused much variation. So, for a few seconds after landing, there must have been some spin around the $X$-axis, or the probe was rolling around the Z-axis, or both. Because we do not know the exact orientation of the probe around landing, we cannot model the variations of the AGC. Even if we could, we would not be able to distinguish between spinning and rolling. Based on the size of the AGC variations after landing compared to before, one can estimate that the spin after landing was on the order of $10^\circ$, or that the rolling after landing was on the order of a few degrees. The roll angle at rest was $0.9^\circ\pm0.5^\circ$ according to \citet{K07}. Thus, the real motion had a significant component out of the $X$-$Z$ plane, but we do not have sufficient information to constrain it.

\subsection{Huygens drop test}
\label{sec:SM2_test}

\begin{figure}
\centering
\includegraphics[width=12cm,angle=0]{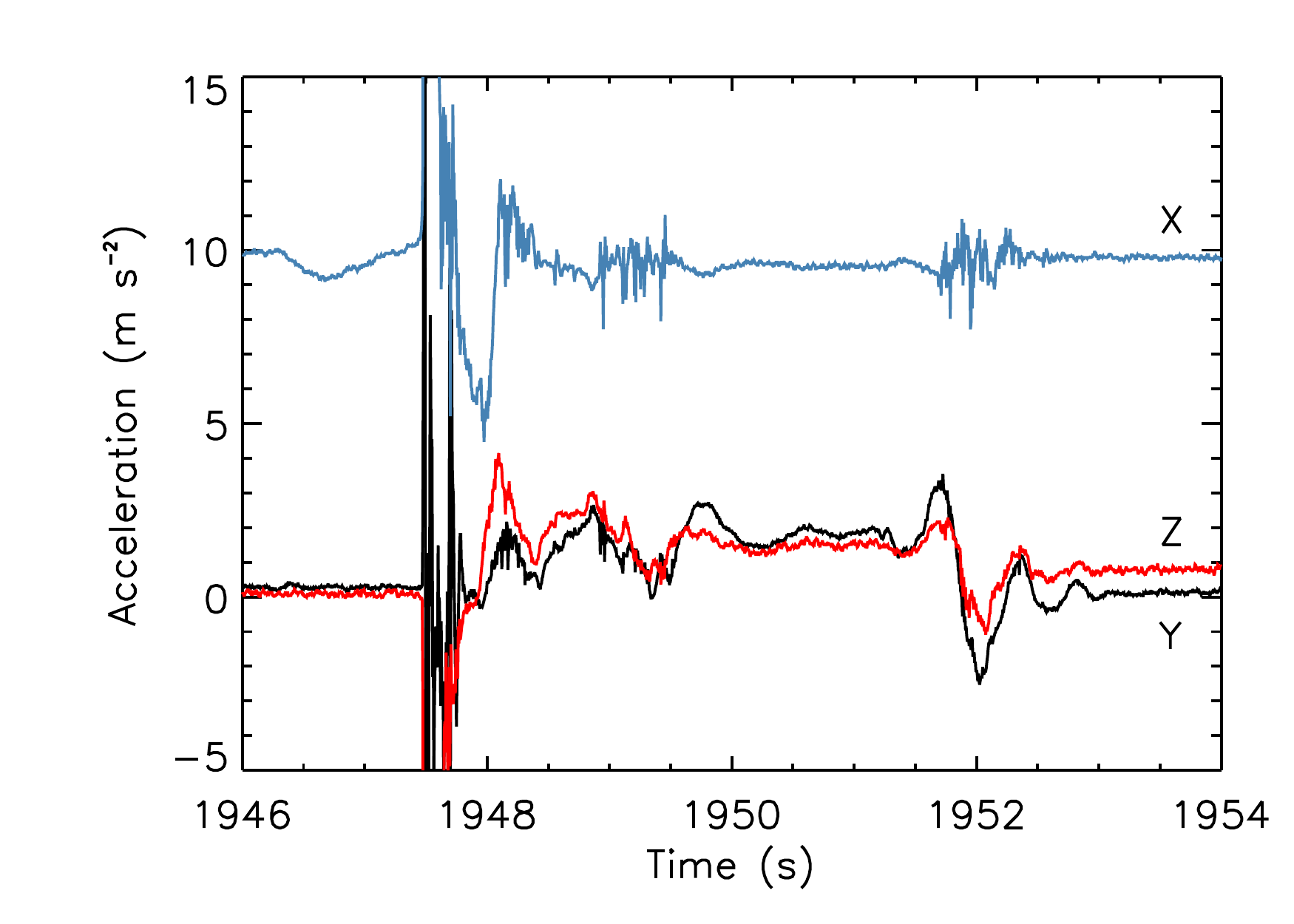}
\includegraphics[width=10cm,angle=0]{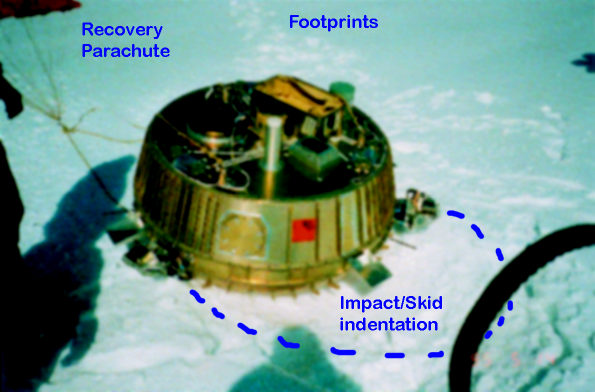}
\caption{The Huygens SM2 model was dropped from a balloon as part of a test in Sweden in 1995. The plot shows the accelerations it experienced at impact; $X$ is the vertical direction, $Y$ and $Z$ are lateral directions. The photo was taken after landing and shows impact markings in the snow, annotated for clarity (photo: ESA).}
\label{fig:SM2}
\end{figure}

Since post-impact survival of Huygens was not considered a design driver for the probe, there was no program of full-scale impact tests. However, some insights can be gleaned from a parachute system test with Special Model~2 (SM2)\footnote{SM2 is now the Huygens model most often seen at exhibitions, and is easily identified by the transparent window on its upper surface used to film the parachute deployment.}, made in 1995 from a stratospheric balloon over Kiruna in Sweden \citep{J96,U97,L10}. Demonstrating the heat shield separation and parachute deployment sequence at flight-like conditions, the SM2 test used a large (non-flight) recovery parachute to accommodate Earth's atmosphere and gravity, which reduced the impact speed to around 8~m~s$^{-1}$. The impact was recorded by onboard accelerometers (Fig.~\ref{fig:SM2}), which clearly show a structural or dynamic bounce about 0.2~s after impact (the vertical acceleration dropped to 5~m~s$^{-2}$ or $0.5g$). Most relevant here, after impact the lateral accelerometers showed acceleration variations on the order of 0.15~m~s$^{-2}$ for a duration of 4~s.

The test report documents the conditions under which the vehicle was recovered \citep{B95}: {\em ``The snow at the landing site was probably 0.5-1~m thick. The descent module had landed on the fore dome and was dragged over about 1m through the snow. The impression in the snow was about 20 to 30~cm deep. The recovery parachute was lying nicely in a straight line next to the descent module. On first sight hardly any damage was visible, except for bending of the spin vanes. Some flattening of the fore dome had occurred. At arrival it could be heard that the gyroscopes were still running.''} An inspection of the photo in the report indicates a smooth depression or skid mark about 0.5~m long (Fig.~\ref{fig:SM2}). Thus in many ways, the landing of the SM2 vehicle resembled the dynamics we have reconstructed for the Huygens probe on Titan.

\section{A Model for Huygens' Motion}
\label{sec:motion_model}

\subsection{Model description}

We have created a description of Huygens' motion after impact that is consistent with both the HASI accelerometer and DISR photometry data sets. It consists of several phases: creation of a hole through impact, bounce out of the hole, slide over the surface, and wobble on the surface. We model the probe motion from the bounce phase to its final resting position. The model is described below, with details of the numerical integration provided in the next section. We use the following basic parameters for Huygens from \citet{L05}: a mass of 200.5~kg, centered about 30~cm above the bottom of the probe, a bottom foredome with a nadir radius of curvature of 1.215~m, and a moment of inertia of 24.5~kg~m$^2$ for rotations around the $Y$-axis. The model is two-dimensional (horizontal and vertical) since the HASI data suggest that most of the motion occurred in the $X$-$Z$ plane.

We start our modeling with the assumption that the surface of Titan was horizontal and flat until Huygens created a 12~cm deep hole upon impact. From the shape of the foredome we infer that this hole had a radius of 53~cm. The probe must have been approximately level at this time because of the smallness of $Y$- and $Z$-accelerations during the impact phase. After bouncing out of the hole the probe slid over the surface, decelerating through friction. The initial phase of the slide is complicated, as the probe did not necessarily end the bounce phase with a purely horizontal motion at ground level. This led to a secondary impact, which is considered in our modeling. We assume a constant friction coefficient throughout the slide. Note that this is the total friction coefficient of the slide, which is not necessarily equal to the friction coefficient of the surface (see Sec.~\ref{sec:friction}). The plateau of the HASI Z-acceleration between mission times 8870.4 and 8871.6~s suggests that the slide lasted about 1.2~s. Integrating this plateau value near $-0.8$~m~s$^{-2}$ over the slide time, or integrating all $Z$-data after the bounce phase, yields a speed near $-1$~m~s$^{-1}$ in both cases. \citet{B08} interpreted this as a horizontal impact speed of 1~m~s$^{-1}$, decreasing to zero after the slide. We refine this estimate by considering that the probe was probably tilted, which means that part of the $Z$-data reflects a component of the gravity, and not a decelerated motion. A tilt is expected on basis of the following argument. The friction force in the negative $Z$-direction acted on the bottom of the probe and thus caused a negative angular acceleration ($\alpha_Y$) of the probe around the $Y$-axis. The front end of the probe would have tilted downward, which we define as a positive pitch angle ($\theta$). Because of the shape of the bottom of the probe, a positive pitch angle causes the contact point with the surface to move ahead of the probe's center of mass, which then causes a positive angular acceleration around the $Y$-axis. Both angular accelerations cancel each other for a pitch angle around $12^\circ$. However, a $12^\circ$ pitch angle causes a $-0.28$~m~s$^{-2}$ component of the gravity vector in the $Z$-direction. Thus, the actual deceleration was somewhat less than the $Z$-plateau value. Also, the actual horizontal impact speed was somewhat less than 1~m~s$^{-1}$, and the equilibrium pitch angle was more like $9^\circ$, consistent with our simulation results.

At the end of the slide, the friction force due to sliding suddenly dropped to zero, which caused a drop of the measured horizontal acceleration. In Fig.~\ref{fig:HASI} we see how the $Z$-acceleration rapidly returned to values near zero at mission time 8871.8~s, which makes this the most likely time when the sliding motion stopped. As we have shown, the pitch angle was now positive and the probe out of balance. This is the start of the wobble phase. For a frictionless wobble, the period of this almost harmonic oscillation would have been 2.2~s, as calculated from Huygens' moment of inertia. In reality, the wobble was damped, and we assume an exponential decay of the amplitude. The $Z$-acceleration (Fig.~\ref{fig:HASI}, top panel) suggests such a decaying wobble, but the period seems to be closer to 1.9-2.0~s. The phasing of the five main DISR data points (bottom panel) is also more consistent with a period near 2.0 rather than 2.2~s. The period is very sensitive to the shape of the surface. A slightly convex surface would increase the period, a concave surface would decrease it. The period near 2.0~s suggests that the surface where the probe came to rest was slightly concave, and we calculate a radius of curvature of 6.5~m. However, DISR post-landing images show a very flat surface with an abundance of pebbles \citep{K07}. One of these sticking out of the surface by 1-2~cm could also have cause the decreased period of wobbling. Additional evidence for a pebble stuck under the probe comes from the pitch angle of $-3.1^\circ\pm0.5^\circ$ calculated for the stable position, which implies a slope of $2.3^\circ\pm0.4^\circ$ for a flat surface \citep{K07}. The time constant for the exponential decay of the wobble amplitude (i.e.\ the time for the amplitude to decrease by a factor of $1/e$) is estimated from the HASI $Z$-acceleration data to be 1.2-1.6~s. The longer time coverage of the DISR data suggests a time constant of 1.4-1.6~s in case the observations roughly probed the peak amplitudes. Thus, we adopt a time constant of 1.5~s. This means that every successive peak amplitude is about half a large as the previous one. In order to model the wobbling motion with decaying amplitude, we assume that the contact point, and thus the location of the force from the ground, is not exactly at the point indicated by the ideal geometry but slightly forward, and that the displacement is proportional to the wobbling speed. The last DISR observation that indicated a position off from the resting position was recorded 7.7~s after impact. At this time, the wobbling amplitude was small, with the pitch angle being only $0.1^\circ$ off the stable value. SSP data indicate continued wobbling motion for another 2~s with even smaller amplitudes.

\subsection{Numerical integration}
\label{sec:numerical}

Our two-dimensional numerical model has a total of six time-dependent variables: two for the location of the probe ($X$ and $Z$), one for the pitch angle ($\theta$), two for both speed components ($v_X$ and $v_Z$), and one for the angular speed around the $Y$-axis ($\omega_Y$). From these we compute the two acceleration components ($a_X$ and $a_Z$) and the angular acceleration ($\alpha_Y$). The accelerations enable us to compute new values for each of the variables for the next time step $\Delta t$, typically chosen as 1 or 2~ms. Model time $t = 0$ is defined at the start of the bounce phase, when Huygens was resting on the bottom of a 12~cm deep hole. For the subsequent bounce, the only force considered is Titan gravity acting on the center of mass (small forces, such as wind resistance, are neglected). Once the probe reaches contact with the surface again, the two components of the surface force are two unknown variables. We distinguish the force components parallel and perpendicular to the motion vector ($F_\parallel$ and $F_\perp$). These are solved with two equations. One equation comes from our assumption that the probe stays constantly in contact with the surface. The other equation is different for sliding and wobbling motion. For the slide, the friction coefficient of the surface ($\mu$) comes into play: $F_\parallel = \mu F_\perp$. For the wobble, the speed is zero for the part of the foredome that is in contact with the surface.

We performed several numerical simulations in which we varied the initial values for the two components of the initial speed and the friction coefficient. For any assumed horizontal speed $v_Z^0$, the vertical speed $v_X^0$ and the friction coefficient $\mu$ can be constrained well from the timing of the HASI acceleration data (Fig.~\ref{fig:HASI}, top panel). If $v_X^0$ is too small or too large, the bounce phase will last too short or too long, respectively, compared to the observed 0.4~s duration. If $\mu$ is too small or too large, the change of the horizontal acceleration $a_Z$ from negative to positive values will occur too late or too early, respectively, compared to the observed mission time of 8871.8~s. We estimate $v_Z^0$ from the HASI $a_Z$ data, and find a best fit for $v_Z^0 = 0.8$-0.9~m~s$^{-1}$. Our result is consistent with the 1~m~s$^{-1}$ from \citet{B08}, who found this value by integrating $a_Z$ on the assumption that the probe did not tilt. For speeds less than 0.6~m~s$^{-1}$, the probe is not able to make it out of its hole during the bounce phase, or it only leaves the hole temporarily, but then rolls back into it. This is inconsistent with the analysis by \citet{K07} indicating that the bottom of the probe was within about 2 cm of to ground level by the time the probe came to rest. Thus, our best guess for the initial horizontal speed is $v_Z^0 = 0.8\pm0.1$~m~s$^{-1}$. Our best guess for the initial vertical speed is $v_X^0 = 0.56$~m~s$^{-1}$. A slightly smaller or larger $v_X^0$ will cause a shorter or longer bounce phase, respectively, than the observed 0.4~s. Thus, $v_X^0$ is better constrained than $v_Z^0$.

The distance covered during the slide over the surface is not well constrained. In the nominal model, the probe comes to rest 86~cm south of the impact point. The resting point is located at least 70~cm south of the impact point (or 17~cm south of the edge of the hole). The 12~cm hole the probe created by the impact has a 53~cm horizontal radius (give or take a few cm due to the uncertainty in depth). Thus the probe cannot end up less than 53 cm south of the impact point because otherwise it would have rolled back into the hole. For a distance between 53 and 70~cm the probe would have gone to the edge of the impact hole during the first wobble back. This would have lengthened the time for that wobble, which is not supported by the data in Fig.~\ref{fig:HASI}. For distances beyond 110~cm south of the impact point, the model fit to the HASI acceleration is not satisfactory. Thus, the slide distance is somewhere between 20 and 50~cm, with a nominal value of 33~cm. The probe slide is terminated by surface friction. We calculated the friction coefficient for the slide as the friction force divided by the weight of the probe. We found reasonable fits for values between 0.3 and 0.5, and adopt $0.4\pm0.1$.

Even though we started our modeling by assuming that the probe was approximately level at the start of the bounce phase, the model fit improves when we allow for a non-zero pitch angle upon impact. This would not affect the shape of the hole, nor would it change the horizontal speed. However, it would make the center of the hole, and thus the impact force, horizontally offset from the probe center of mass. This, in turn, would cause the probe to leave the hole with a non-zero angular speed. Our adopted model has an initial angular speed of $\omega_Y^0 = 0.15$~s$^{-1}$, but this value is not well constrained. Considering the change of the vertical speed of $\Delta v_X = 5.1$~m~s$^{-1}$ and the geometry of the probe, this $\omega_Y^0$ value is obtained in case the center of mass is offset by 0.4~cm, corresponding to a pitch angle at impact of $-0.23^\circ$. This value is well within the average tilt angle of $1.5^\circ$ obtained for the descent at low altitudes \citep{K07}. A negative pitch angle is also consistent with negative $Z$-accelerations recorded before impact. Furthermore, as shown in Sec.~\ref{sec:wind}, the independently inferred vertical wind speed profile would also make the parachute-probe system tilt to negative pitch angles seconds before impact.

\begin{figure}
\centering
\includegraphics[width=14cm,angle=-90]{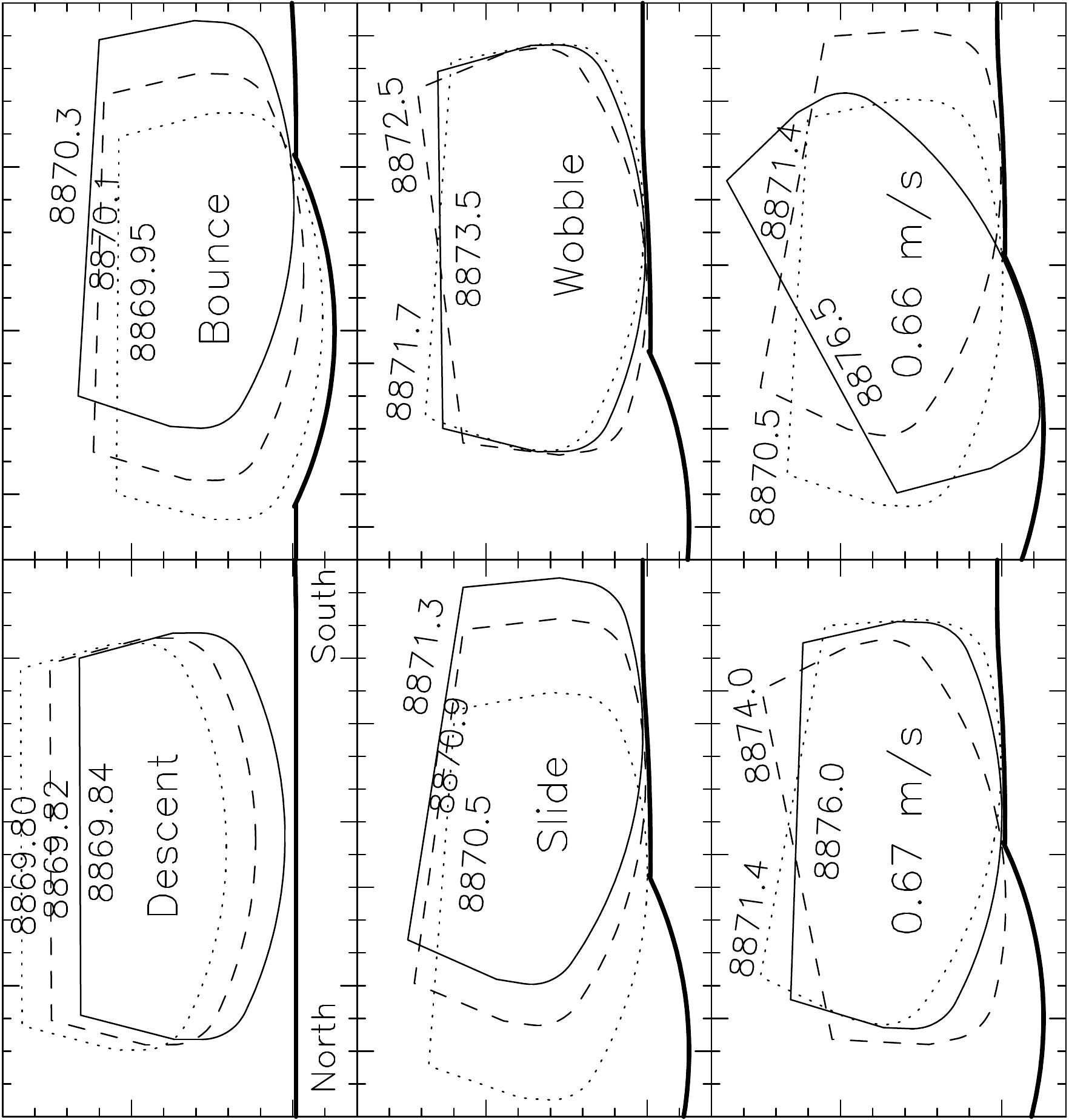}
\caption{The four phases around landing with three locations of the probe for each case and the last phase for two alternative initial horizontal speeds. In each panel, the location of the probe for three mission times is shown as indicated, in the temporal order of dotted, dashed, and solid outline. The bottom two panels show the inferred motion if the horizontal component of the impact speed would have been $v_Z^0 = 0.67$ or 0.66~m~s$^{-1}$ instead of 0.80~m~s$^{-1}$. Small tick marks are spaced by 10~cm, large tick marks by 50~cm.}
\label{fig:Huygens_motion}
\end{figure}

Figure~\ref{fig:Huygens_motion} shows the four main phases around the time of landing. A movie shows the motion and the forces (supplementary material). At the bottom two panels of Fig.~\ref{fig:Huygens_motion} and for two additional movies, we also considered cases of horizontal speeds of $v_Z^0 = 0.67$ and 0.66~m~s$^{-1}$, instead of the adopted 0.80~m~s$^{-1}$, leaving all other parameters unchanged. In the case of $v_Z^0 = 0.67$~m~s$^{-1}$, the probe takes a very long time for its first wobble period, which is inconsistent with observations. In the case of $v_Z^0 = 0.66$~m~s$^{-1}$, the probe rolls back into its hole and ends up with a pitch angle $\theta = -29^\circ$, very inconsistent with observations.

\subsection{Rolling over a pebble}

When comparing the predictions of our motion model with the HASI accelerations in Fig.~\ref{fig:HASI} (top panel) we note two major deviations, one between mission times 8871.5 and 8872.0~s, and the other between 8872.0 and 8872.8~s. First we focus on the latter event.

Had Huygens rolled over a pebble, it would have left a clear signature in the acceleration data. The recorded vertical acceleration would suddenly jump way above Titan's gravity, as the vertical speed changes to the positive value needed to climb the pebble. Then, during the roll over the pebble, the vertical speed would drop to negative values, which decreases the recorded vertical acceleration to values below Titan's gravity. Finally, when the ground is reached again, the recorded acceleration jumps for a short time above Titan's gravity. In case the probe stopped on top of the pebble and rolled back where it came from, the recorded accelerations would have been similar. Considering that rolling friction slowed the probe down, the initial motion onto the pebble would have caused larger accelerations, of shorter duration, compared to the exit from the pebble. There is one event in the HASI data that matches the expected accelerations for rolling over a pebble almost perfectly, between mission times 8872.0 and 8872.8~s. During this period, the rolling motion reverses according to our model. We integrated the offset between the observed acceleration and the modeled values twice, and we determined the two integration constants by setting the height of the probe to zero at the beginning and end of this period. Then, the height of the probe in the middle of this period comes out to 2~cm. Thus, the probe rolled over a pebble sticking 2~cm out of the ground. Since this took place when the probe was tilted with a negative pitch angle, we find a similar signature with same sign, but lower amplitude, in the $Z$-acceleration data. If Huygens rolled over a pebble during the first wobble back, one might expect it should have gone over the same pebble during the slide. Yet, the data suggest this was not the case. Our expectation is only true if the motion was purely in the $X$-$Z$ plane. Likely, some smaller motion also occurred in the $Y$-direction. Thus, Huygens probably did not exactly trace back the track of the slide during its first wobble back. This explains why it only encountered the pebble during the first wobble back, but not during the slide.

Our motion model cannot reproduce the other major deviation, the $X$-accelerations between 8871.5 and 8872.0~s. All model runs create values near Titan's gravity while the data show only half that value. The only way to match these data is to start with a larger initial vertical speed for the bounce which puts the probe about 50~cm above ground for the slide and wobble. Since the DISR images do not show such variations of the terrain, we consider this unlikely. The analysis by \citet{B08} and the difference between the impact record of the HASI and SSP accelerometers show that structural oscillations of the experiment platform significantly affect the measured accelerations during the first second after impact, which may explain the discrepancy.

\subsection{Friction coefficient}
\label{sec:friction}

During the slide on the surface, Huygens decelerated through friction. Our model calculates a friction coefficient for this slide, with a nominal value of $\mu_{\rm T} = 0.4\pm0.1$ (Sec.~\ref{sec:numerical}). However, this is not simply the friction coefficient of the surface proper. The situation is complicated by the structures protruding from the probe foredome. One of these is the SSP penetrometer, which stuck out by 5.5~cm, somewhat offset in the $Y$-direction. During the slide, it would have dug a 1.5~cm wide trench. The depth of the trench depends on the tilt in both directions. With an average pitch angle of $8^\circ$, the trench would have been 2-3~cm deep. The friction caused by digging the trench may have been significant. The total (model) friction force experienced during the slide is the sum of the force due to the penetrometers and the classical surface friction force for a smooth slide: $F_\parallel^{\rm T} = F_\parallel^{\rm P} + F_\parallel^{\rm S}$. We find an identical expression for the friction coefficient: $\mu_{\rm T} = \mu_{\rm P} + \mu_{\rm S}$. Force and friction coefficient are related through $F_\parallel = \mu F_\perp$, with $F_\perp = 268$~N, the probe weight on Titan. The quantity of physical interest is the friction coefficient of the surface material $\mu_{\rm S}$, which, given $\mu_{\rm T}$ from our model, we can only calculate if we have a good estimate for $\mu_{\rm P}$. The side-force on the penetrometer ($F_\parallel^{\rm P}$) can be estimated as follows. The size of the penetrometer is around 5~cm $\times$ 1~cm. The bearing strength of the surface indicated by the penetrometer was around 50~N~cm$^{-2}$ (500~kPa) over an area of 1-2~cm$^2$ \citep{Z05,A10}, whereas the deceleration of the 205~kg probe ($18g$) implies a peak force of 36~kN over about 1~m$^2$ or less of probe area, and thus 50~kPa or 5~N~cm$^{-2}$ \citep{L09}. Adopting these two values as extremes, the force on the penetrometer might have been 25-250~N. Similar-sized protrusions are represented by the GCMS and ACP inlets near the apex of the probe, so the side-force could plausibly have been 2-4 times larger, i.e.\ a force range of $F_\parallel^{\rm P} = 50$-1000~N. The effective friction coefficient due to the protrusions ($\mu_{\rm P}$), ignoring any skin friction, could therefore be as low as 0.2 or greater than unity.

Thus, it seems impossible to derive a good estimate for $\mu_{\rm S}$, the friction coefficient of the surface material itself. The dynamics suggest that the material was not a viscous or plastic mud, as this would simply have let the vehicle stop with a splat. The material had to have an elastic component in its rheology. Dry or damp sand would work, although dry sand appears to be excluded by other data (e.g. \citealt{L09}).

\subsection{Surface wind}
\label{sec:wind}

Our motion model can constrain the wind velocity at the surface. The velocity near the surface was inferred by \citet{K07} by tracking the parachute as it partially obstructed the solar aureole 9-14~s after impact (mission time 8879-8884~s)\footnote{\citet{K07} detected obstruction of the solar aureole by the parachute in the DISR ULIS data. We looked for a similar signature in ULV data, but did not find any. We estimate the effect of parachute obstruction for the ULV to be roughly 3\%, corresponding to 0.2~DN. One ULV measurement at 8883~s (\#460) could be slightly affected, but a 0.2~DN decrease is too low to make a significant difference.}. Assuming that the parachute followed the wind, the wind speed was $0.3\pm0.1$~m~s$^{-1}$ toward azimuth $160^\circ\pm10^\circ$ (roughly SSE) around 5-10~m above the ground. However, this result assumes no probe motion after impact. If we take our model with a 86~cm southward motion between impact and coming to rest, the parachute speed changes slightly to $0.4\pm0.1$~m~s$^{-1}$ in the direction of azimuth $165^\circ\pm10^\circ$.

A second estimate of the wind speed was obtained by \citet{L06}, and is based on the observed slow cooling of the probe by wind during its time on the surface. He derived an upper limit for the average wind speed of 0.25~m~s$^{-1}$ within 1~m of the ground level. Our model provides a third estimate. The probe had a horizontal impact speed component of 0.8~m~s$^{-1}$ toward azimuth $190^\circ$. Considering the length of the parachute (12~m) and the delayed action from the parachute to the probe, this horizontal motion reflects the wind speed at 20-30~m altitude. These three results suggest that the wind direction near the ground was close to north-south, while the wind speed decreased toward the ground in the lowest 30~m. If true, then the decreasing wind decelerated the parachute during the last seconds before impact. A decelerating parachute tilts the lines between it and the probe, so that the probe is slightly forward (south) of the parachute. Our model concluded such a tilt at impact based on the data after impact.

\begin{figure}
\centering
\includegraphics[angle=-90,width=\textwidth]{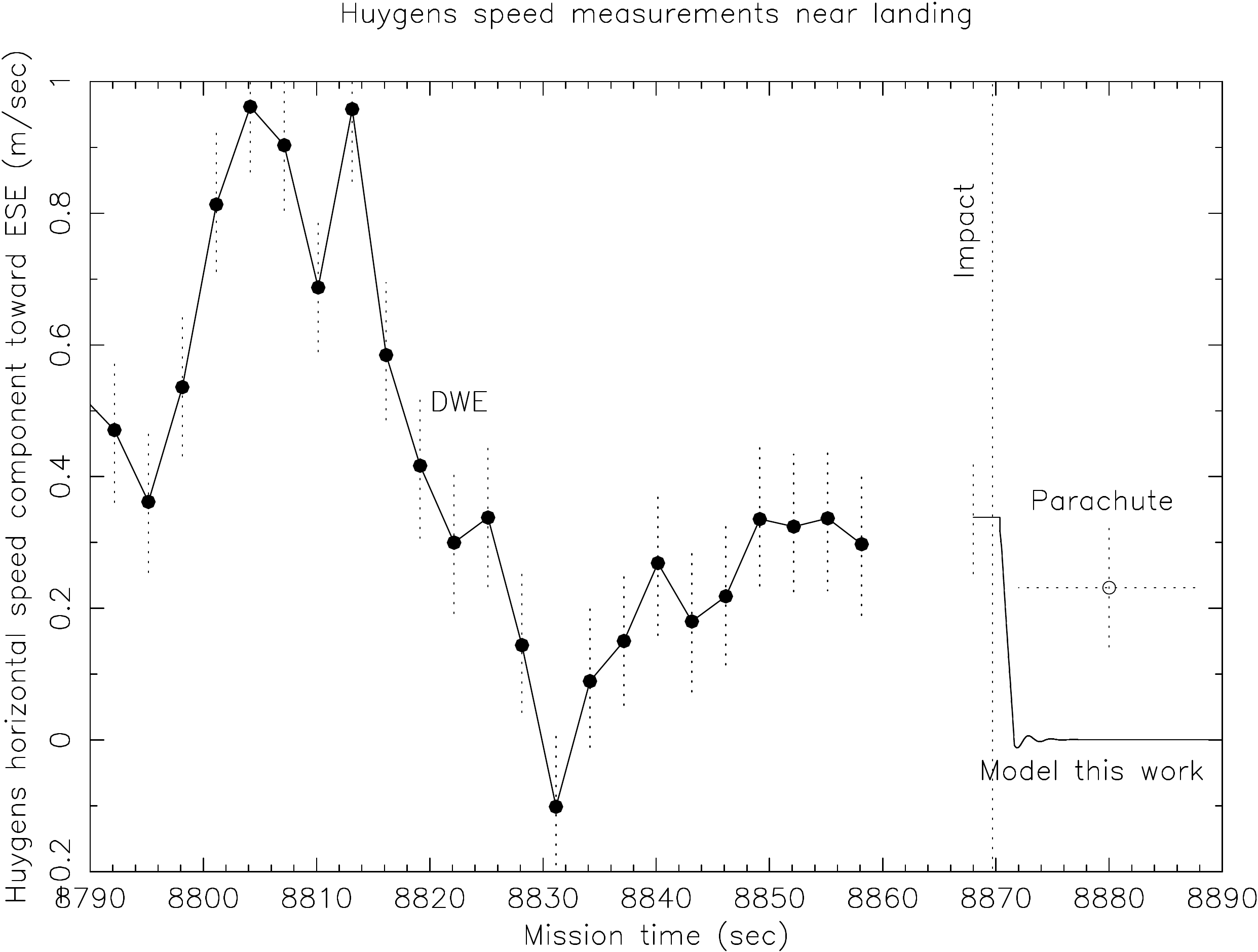}
\caption{Horizontal speed data for the component in direction azimuth $115^\circ$ (roughly ESE) around the time of landing from DWE data (left), DLIS data of the obstruction of the parachute (right), and from our model (bottom right). Note that the parachute data point applies for the motion of the parachute, not the probe. Error bars are indicated by dotted lines. The last 80~s before impact shown here correspond to the descent from 360~m altitude to the surface (calculated from an average descent speed of 4.5~m~s$^{-1}$).}
\label{fig:DWE}
\end{figure}

Figure~\ref{fig:DWE} shows how our findings compare with wind speeds measured by the Doppler Wind Experiment (DWE) aboard Huygens \citep{B05}. The DWE gives us only the zonal motion, i.e.\ the horizontal speed profile in the direction of azimuth $115^\circ$ (ESE). The speed in the perpendicular direction (SSW) is unconstrained. During the last minute before impact, the probe speed varied by at least 1~m~s$^{-1}$, possibly more considering that Fig.~\ref{fig:DWE} shows only one component of the speed. The wind speed probably varied even more than that, as the probe speed only partially follows wind variations on time scales less than 10~s. Our model impact speed component aligns quite well with the last measurements of DWE before impact (Fig.~\ref{fig:DWE}).

\section{Discussion}
\label{sec:discussion}

We have analyzed data from a wide variety of instruments onboard Huygens and conclude that all show evidence for continued movement for several seconds after impact on the surface of Titan. We propose the following scenario. Upon impact Huygens created a 12~cm deep hole. It bounced out of the hole 0.4~s after impact. Its horizontal speed was sufficient to move it out onto the almost flat surface, where it commenced a slide. Due to the initial angular speed and the friction force, the probe tilted to a positive pitch angle, i.e.\ the front end of its motion, where DISR was located, was closer to the ground by almost 10~cm. The slide covered a distance of 30-40~cm in 1.2~s. At the end of the slide, the friction force due to the motion disappeared, the probe was out of balance and started to wobble. At this point, it encountered a pebble sticking out 2~cm above the ground. The measurements indicate that the probe wobbled back and fourth five times with decreasing amplitude. The last DISR observation indicating movement was acquired 7.7~s after impact, consistent with a displacement of 2~mm, which is well above our detection threshold of 0.5~mm. SSP data suggest the last measurable excursion from the stable position was about 10~s after impact, with a tilt offset from the stable position of about $0.02^\circ$. Similarities observed during a full-scale Huygens drop test support this scenario.

We simulated this motion with a numerical model using a total of seven parameters: three for the initial conditions immediately after impact, two for the friction of a sliding and wobbling probe, and two for the topography of the terrain. We found a set of parameters (the ``nominal model'') that explains all main features of post-impact DISR photometric observations and HASI acceleration data\footnote{The aforementioned uncertainty regarding the HASI $Y$-axis orientation has only minor consequences for the outcome of our simulations. It would only change the direction of the slide from $3^\circ$ left of the $Z$-axis to $3^\circ$ to the right of the $Z$-axis, while the wobble from the SSP remains at $3^\circ$ to the left of the $Z$-axis. This is certainly possible, since the probe could have rotated a few degrees or the motion might not have been perfectly linear.}: an initial vertical speed of 0.56~m~s$^{-1}$, horizontal speed of 0.8~m~s$^{-1}$ in the $Z$-direction pointing roughly south, angular speed of 0.15~s$^{-1}$ around the $Y$-axis, friction coefficient for the sliding motion of 0.4, damping time scale of the wobbling motion of 1.5~s, a concave shape of the probed surface with a radius of curvature of 6.5~m, and a slope of the surface moving up to the south of $2.3^\circ$ at the resting location 86~cm south of the initial impact point. An alternative explanation for the concave shape and slope of the surface is the presence of a pebble under the probe.

What does our modeling effort tell us about the physical properties of Titan's surface at the landing site? In principle, the friction coefficient for the slide of 0.4 puts constraints on the surface material, but an accurate consideration needs to account for the shape of the foredome including various extrusions, which is beyond the scope of this work (and would most likely be too poorly constrained to be useful). Huygens' wobble spanning five full periods indicates that the surface was able to support the probe weight without much deformation. On the other hand, the formation of a 12~cm deep hole upon impact implies a relatively soft surface. These considerations certainly put constraints on the properties of Titan's surface, but are not readily reconciled. By comparing SSP measurements with laboratory experiments, \citet{A10} identified the most likely candidate for the surface at the landing site to be damp and cohesive material with interstitial liquid contained between its grains, a notion also supported by GCMS results \citep{L06b}. Perhaps the force of the impact promoted the formation of the hole by means of liquefaction, the process by which saturated, unconsolidated sediment is transformed into a substance that acts like a liquid. Note that the surface at the landing site is thought to be sedimentary in nature \citep{So07}. The secondary impact after the bounce was much less energetic and would have left the soil firm. Further laboratory experiments would help to assess the likelihood of this scenario.

We attribute observed changes in the slope of DLVS spectra acquired in the first 4~s after impact to a transient dust cloud. Landers on other planets have also created transient dust clouds: several Venus Venera landers detected dust \citep{M79,G82}, as did the Mars Viking landers \citep{M76}. It would result from the deposition of the turbulent wake behind the probe on the surface \citep{L93,E00}. There is evidence for a 7~mm thin layer with mechanical properties similar to terrestrial snow on the surface at the landing site \citep{A10}. It is only natural to assume that this layer is composed of organic aerosols that form continuously in the global haze layer, and settle on the surface of Titan at a rate of around 4~mg~m$^{-2}$~yr$^{-1}$ \citep{McK89}. The observed spectral changes are modeled well on the assumption that the dust has optical properties identical to those derived for the aerosols \citep{T08}. The fact that dust was lofted into the air implies the absence of strong cohesive forces between the particles. This is consistent with the idea that, despite evidence for cohesion and moisture in the very shallow ($\sim$cm) subsurface \citep{N05,L06b,A10}, the surface itself was not wet. How can this be reconciled with the identification of a drop of liquid methane in one of the DISR images by \citet{KT09}? The drop is thought to have condensed on the cold baffle of the Side Looking Imager, which presumes evaporation from the (sub)surface. \citet{L06} showed that the power density of the probe heat leak was much smaller than that of the lamp-heated spot. With dust blown
away by impact, methane may have evaporated from the lamp spot. However, the DLIS did not observe any increase in atmospheric methane abundance after landing, while peering directly into the warm lamp reflection spot \citep{SK08}. While we are unable to reconcile all data, we tentatively conclude the detection of dust on the surface. Assuming the liquid came from above, a dry surface implies that since the wetting event there had been a dry period that desiccated several millimeters of surface material. This is consistent with the paradigm that Titan's hydrologic cycle sees long droughts between rainstorms \citep{L00}.

\section{Conclusions}
\label{sec:conclusions}

Data acquired by Huygens probe support the notion that the probe moved for several seconds after landing on the surface of Titan. Our motion model explains all major features present in HASI accelerometer data, and is consistent with DISR photometric measurements. Upon impact, Huygens created a 12~cm deep hole from which it bounced back onto the surface. It then performed a 30-40~cm long slide, after which it wobbled back and forth five times until it finally came to rest around 10~s after impact. We conclude that the surface was soft enough to permit the formation of a hole, and hard enough to support the probe wobble. Unfortunately, the presence of protruding structures at the underside of the probe does not allow us to derive a meaningful estimate for the friction coefficient of the surface. Spectral variability detected by DISR right after impact is consistent with a transient dust cloud formed by the impact of the turbulent wake behind the probe on the surface. The dust was most likely composed of organic aerosols that continuously settle on the surface of Titan, and its presence implies a recent dry period.

\section*{Acknowledgements}

SES is grateful for the generous support provided by Marty Tomasko and his team at the LPL in Tucson, Arizona. RDL acknowledges the support of the Cassini project, and thanks Mark Leese and Brijen Hathi for re-examining the timing of the SSP tilt measurements. We thank Chuck See and Lyn Doose for providing DISR calibration data and their suggestions for improving the manuscript. In addition, we thank two anonymous referees for their helpful comments. Support for this work was provided by the {\em Deutsches Zentrum f\"ur Luft und Raumfahrt} (DLR) through grant 50 OH 98044 and by NASA through grants NNX10AF09G (EK) and NNX11AC98G (RDL). After this paper was accepted, C. Bettanini kindly informed us that the HASI piezo accelerometer was installed along the Huygens $Y$-axis, but with the sensing direction in the $-Y$ and not the $+Y$ direction.



\bibliography{around_landing}

\begin{thebibliography}{38}
\expandafter\ifx\csname natexlab\endcsname\relax\def\natexlab#1{#1}\fi
\expandafter\ifx\csname url\endcsname\relax
  \def\url#1{\texttt{#1}}\fi
\expandafter\ifx\csname urlprefix\endcsname\relax\def\urlprefix{URL }\fi

\bibitem[{{Atkinson} et~al.(2010){Atkinson}, {Zarnecki}, {Towner}, {Ringrose},
  {Hagermann}, {Ball}, {Leese}, {Kargl}, {Paton}, {Lorenz}, and {Green}}]{A10}
{Atkinson}, K.~R., {Zarnecki}, J.~C., {Towner}, M.~C., {Ringrose}, T.~J.,
  {Hagermann}, A., {Ball}, A.~J., {Leese}, M.~R., {Kargl}, G., {Paton}, M.~D.,
  {Lorenz}, R.~D., {Green}, S.~F., Dec. 2010. {Penetrometry of granular and
  moist planetary surface materials: Application to the Huygens landing site on
  Titan}. \icarus 210, 843--851.

\bibitem[{{Bettanini} et~al.(2008){Bettanini}, {Zaccariotto}, and
  {Angrilli}}]{B08}
{Bettanini}, C., {Zaccariotto}, M., {Angrilli}, F., Apr. 2008. {Analysis of the
  HASI accelerometers data measured during the impact phase of the Huygens
  probe on the surface of Titan by means of a simulation with a finite-element
  model}. \planss 56, 715--727.

\bibitem[{{Bird} et~al.(2005){Bird}, {Allison}, {Asmar}, {Atkinson}, {Avruch},
  {Dutta-Roy}, {Dzierma}, {Edenhofer}, {Folkner}, {Gurvits}, {Johnston},
  {Plettemeier}, {Pogrebenko}, {Preston}, and {Tyler}}]{B05}
{Bird}, M.~K., {Allison}, M., {Asmar}, S.~W., {Atkinson}, D.~H., {Avruch},
  I.~M., {Dutta-Roy}, R., {Dzierma}, Y., {Edenhofer}, P., {Folkner}, W.~M.,
  {Gurvits}, L.~I., {Johnston}, D.~V., {Plettemeier}, D., {Pogrebenko}, S.~V.,
  {Preston}, R.~A., {Tyler}, G.~L., Dec. 2005. {The vertical profile of winds
  on Titan}. \nat 438, 800--802.

\bibitem[{{Brouwer}(1995)}]{B95}
{Brouwer}, G.~F., 1995. {Huygens SM2 balloon air drop test: Inspection report
  of recovered items}. Tech. Rep. HUY-FOKK-532-RE-0031, Fokker Space \& Systems
  B.V., Leiden, The Netherlands.

\bibitem[{{Eames} and {Dalziel}(2000)}]{E00}
{Eames}, I., {Dalziel}, S.~B., Jan. 2000. {Dust resuspension by the flow around
  an impacting sphere}. Journal of Fluid Mechanics 403, 305--328.

\bibitem[{{Fulchignoni} et~al.(2002){Fulchignoni}, {Ferri}, {Angrilli},
  {Bar-Nun}, {Barucci}, {Bianchini}, {Borucki}, {Coradini}, {Coustenis},
  {Falkner}, {Flamini}, {Grard}, {Hamelin}, {Harri}, {Leppelmeier},
  {Lopez-Moreno}, {McDonnell}, {McKay}, {Neubauer}, {Pedersen}, {Picardi},
  {Pirronello}, {Rodrigo}, {Schwingenschuh}, {Seiff}, {Svedhem}, {Vanzani}, and
  {Zarnecki}}]{F02}
{Fulchignoni}, M., {Ferri}, F., {Angrilli}, F., {Bar-Nun}, A., {Barucci},
  M.~A., {Bianchini}, G., {Borucki}, W., {Coradini}, M., {Coustenis}, A.,
  {Falkner}, P., {Flamini}, E., {Grard}, R., {Hamelin}, M., {Harri}, A.~M.,
  {Leppelmeier}, G.~W., {Lopez-Moreno}, J.~J., {McDonnell}, J.~A.~M., {McKay},
  C.~P., {Neubauer}, F.~H., {Pedersen}, A., {Picardi}, G., {Pirronello}, V.,
  {Rodrigo}, R., {Schwingenschuh}, K., {Seiff}, A., {Svedhem}, H., {Vanzani},
  V., {Zarnecki}, J., Jul. 2002. {The Characterisation of Titan's Atmospheric
  Physical Properties by the Huygens Atmospheric Structure Instrument (HASI)}.
  \ssr 104, 395--431.

\bibitem[{{Garvin}(1982)}]{G82}
{Garvin}, J.~B., 1982. {Landing induced dust clouds on Venus and Mars}. In:
  {R.~B.~Merrill \& R.~Ridings} (Ed.), Lunar and Planetary Science Conference
  Proceedings. Vol.~12 of Lunar and Planetary Science Conference Proceedings.
  pp. 1493--1505.

\bibitem[{{Hathi} et~al.(2009){Hathi}, {Ball}, {Colombatti}, {Ferri}, {Leese},
  {Towner}, {Withers}, {Fulchigioni}, and {Zarnecki}}]{H09}
{Hathi}, B., {Ball}, A.~J., {Colombatti}, G., {Ferri}, F., {Leese}, M.~R.,
  {Towner}, M.~C., {Withers}, P., {Fulchigioni}, M., {Zarnecki}, J.~C., Oct.
  2009. {Huygens HASI servo accelerometer: A review and lessons learned}.
  \planss 57, 1321--1333.

\bibitem[{{J{\"a}kel} et~al.(1996){J{\"a}kel}, {Rideau}, {Nugteren}, and
  {Underwood}}]{J96}
{J{\"a}kel}, E., {Rideau}, P., {Nugteren}, P.~R., {Underwood}, J., Feb. 1996.
  {Drop testing the Huygens Probe}. ESA Bulletin 85.

\bibitem[{{Karkoschka} et~al.(2012){Karkoschka}, {Schr{\"o}der}, {Tomasko}, and
  {Keller}}]{K12}
{Karkoschka}, E., {Schr{\"o}der}, S.~E., {Tomasko}, M.~G., {Keller}, H.~U.,
  Jan. 2012. {The reflectivity spectrum and opposition effect of Titan's
  surface observed by Huygens' DISR spectrometers}. \planss 60, 342--355.

\bibitem[{{Karkoschka} and {Tomasko}(2009)}]{KT09}
{Karkoschka}, E., {Tomasko}, M.~G., Feb. 2009. {Rain and dewdrops on titan
  based on in situ imaging}. \icarus 199, 442--448.

\bibitem[{{Karkoschka} et~al.(2007){Karkoschka}, {Tomasko}, {Doose}, {See},
  {McFarlane}, {Schr\"oder}, and {Rizk}}]{K07}
{Karkoschka}, E., {Tomasko}, M.~G., {Doose}, L.~R., {See}, C., {McFarlane},
  E.~A., {Schr\"oder}, S.~E., {Rizk}, B., Nov. 2007. {DISR Imaging and the
  Geometry of the Descent of the Huygens Probe within Titan's Atmosphere}.
  \planss 55, 1896--1935.

\bibitem[{{Lebleu}(2005)}]{L05}
{Lebleu}, D., Jun. 2005. {Huygens Probe: Probe reference data for post flight
  analysis}. Tech. Rep. HUY.ASP.MIS.TN.0006, Alcatel Space, Cannes, France.

\bibitem[{{Lebreton} and {Matson}(2002)}]{L02}
{Lebreton}, J.-P., {Matson}, D.~L., Jul. 2002. {The Huygens Probe: Science,
  Payload and Mission Overview}. \ssr 104, 59--100.

\bibitem[{{Lebreton} et~al.(2005){Lebreton}, {Witasse}, {Sollazzo},
  {Blancquaert}, {Couzin}, {Schipper}, {Jones}, {Matson}, {Gurvits},
  {Atkinson}, {Kazeminejad}, and {P{\'e}rez-Ay{\'u}car}}]{LW05}
{Lebreton}, J.-P., {Witasse}, O., {Sollazzo}, C., {Blancquaert}, T., {Couzin},
  P., {Schipper}, A.-M., {Jones}, J.~B., {Matson}, D.~L., {Gurvits}, L.~I.,
  {Atkinson}, D.~H., {Kazeminejad}, B., {P{\'e}rez-Ay{\'u}car}, M., Dec. 2005.
  {An overview of the descent and landing of the Huygens probe on Titan}. \nat
  438, 758--764.

\bibitem[{{Leese} et~al.(2012){Leese}, {Lorenz}, {Hathi}, and {Zarnecki}}]{L12}
{Leese}, M.~R., {Lorenz}, R.~D., {Hathi}, B., {Zarnecki}, J.~C., Sep. 2012.
  {The Huygens surface science package (SSP): Flight performance review and
  lessons learned}. \planss 70, 28--45.

\bibitem[{{Lorenz}(1993)}]{L93}
{Lorenz}, R.~D., Jan. 1993. {Wake-induced dust cloud formation following impact
  of planetary landers}. \icarus 101, 165--167.

\bibitem[{{Lorenz}(2000)}]{L00}
{Lorenz}, R.~D., Oct. 2000. {The Weather on Titan}. Science 290, 467--468.

\bibitem[{{Lorenz}(2006)}]{L06}
{Lorenz}, R.~D., Jun. 2006. {Thermal interactions of the Huygens probe with the
  Titan environment: Constraint on near-surface wind}. \icarus 182, 559--566.

\bibitem[{{Lorenz}(2010)}]{L10}
{Lorenz}, R.~D., Apr. 2010. {Attitude and angular rates of planetary probes
  during atmospheric descent: Implications for imaging}. \planss 58, 838--846.

\bibitem[{{Lorenz} et~al.(2009){Lorenz}, {Kargl}, {Ball}, {Zarnecki}, {Towner},
  {Leese}, {McDonnell}, {Atkinson}, {Hathi}, and {Hagermann}}]{L09}
{Lorenz}, R.~D., {Kargl}, G., {Ball}, A.~J., {Zarnecki}, J.~C., {Towner},
  M.~C., {Leese}, M.~R., {McDonnell}, J.~A.~M., {Atkinson}, K.~R., {Hathi}, B.,
  {Hagermann}, A., 2009. {Titan surface mechanical properties from the SSP
  ACC-I record of the impact deceleration of the Huygens probe}. In: {G.~Kargl,
  N.~I.~K{\"o}mle, A.~J.~Ball, \& R.~D.~Lorenz} (Ed.), Penetrometry in the
  Solar System II. p. 147.

\bibitem[{{Lorenz} et~al.(2006){Lorenz}, {Niemann}, {Harpold}, {Way}, and
  {Zarnecki}}]{L06b}
{Lorenz}, R.~D., {Niemann}, H.~B., {Harpold}, D.~N., {Way}, S.~H., {Zarnecki},
  J.~C., 2006. {Titan's damp ground: Constraints on Titan surface thermal
  properties from the temperature evolution of the Huygens GCMS inlet}.
  Meteoritics and Planetary Science 41, 1705--1714.

\bibitem[{{Lorenz} et~al.(2007){Lorenz}, {Zarnecki}, {Towner}, {Leese}, {Ball},
  {Hathi}, {Hagermann}, and {Ghafoor}}]{L07}
{Lorenz}, R.~D., {Zarnecki}, J.~C., {Towner}, M.~C., {Leese}, M.~R., {Ball},
  A.~J., {Hathi}, B., {Hagermann}, A., {Ghafoor}, N.~A.~L., Nov. 2007. {Descent
  motions of the Huygens probe as measured by the Surface Science Package
  (SSP): Turbulent evidence for a cloud layer}. \planss 55, 1936--1948.

\bibitem[{{McKay} et~al.(1989){McKay}, {Pollack}, and {Courtin}}]{McK89}
{McKay}, C.~P., {Pollack}, J.~B., {Courtin}, R., Jul. 1989. {The thermal
  structure of Titan's atmosphere}. \icarus 80, 23--53.

\bibitem[{{Moshkin} et~al.(1979){Moshkin}, {Ekonomov}, and {Golovin}}]{M79}
{Moshkin}, B.~E., {Ekonomov}, A.~P., {Golovin}, I.~M., Sep. 1979. {Dust on the
  surface of Venus}. Cosmic Research 17, 232--237.

\bibitem[{{Mutch} et~al.(1976){Mutch}, {Grenander}, {Jones}, {Patterson},
  {Arvidson}, {Guinness}, {Avrin}, {Carlston}, {Binder}, and {Sagan}}]{M76}
{Mutch}, T.~A., {Grenander}, S.~U., {Jones}, K.~L., {Patterson}, W.,
  {Arvidson}, R.~E., {Guinness}, E.~A., {Avrin}, P., {Carlston}, C.~E.,
  {Binder}, A.~B., {Sagan}, C., Dec. 1976. {The surface of Mars - The view from
  the Viking 2 lander}. Science 194, 1277--1283.

\bibitem[{{Niemann} et~al.(2005){Niemann}, {Atreya}, {Bauer}, {Carignan},
  {Demick}, {Frost}, {Gautier}, {Haberman}, {Harpold}, {Hunten}, {Israel},
  {Lunine}, {Kasprzak}, {Owen}, {Paulkovich}, {Raulin}, {Raaen}, and
  {Way}}]{N05}
{Niemann}, H.~B., {Atreya}, S.~K., {Bauer}, S.~J., {Carignan}, G.~R., {Demick},
  J.~E., {Frost}, R.~L., {Gautier}, D., {Haberman}, J.~A., {Harpold}, D.~N.,
  {Hunten}, D.~M., {Israel}, G., {Lunine}, J.~I., {Kasprzak}, W.~T., {Owen},
  T.~C., {Paulkovich}, M., {Raulin}, F., {Raaen}, E., {Way}, S.~H., Dec. 2005.
  {The abundances of constituents of Titan's atmosphere from the GCMS
  instrument on the Huygens probe}. \nat 438, 779--784.

\bibitem[{{Schr\"oder}(2007)}]{S07}
{Schr\"oder}, S.~E., 2007. {Investigating the Surface of Titan with the Descent
  Imager/Spectral Radiometer onboard Huygens}. Ph.D. thesis, {Universit\"at
  G\"ottingen, Germany}.

\bibitem[{{Schr{\"o}der} and {Keller}(2008)}]{SK08}
{Schr{\"o}der}, S.~E., {Keller}, H.~U., Apr. 2008. {The reflectance spectrum of
  Titan's surface at the Huygens landing site determined by the Descent
  Imager/Spectral Radiometer}. \planss 56, 753--769.

\bibitem[{{Soderblom} et~al.(2007){Soderblom}, {Tomasko}, {Archinal}, {Becker},
  {Bushroe}, {Cook}, {Doose}, {Galuszka}, {Hare}, {Howington-Kraus},
  {Karkoschka}, {Kirk}, {Lunine}, {McFarlane}, {Redding}, {Rizk}, {Rosiek},
  {See}, and {Smith}}]{So07}
{Soderblom}, L.~A., {Tomasko}, M.~G., {Archinal}, B.~A., {Becker}, T.~L.,
  {Bushroe}, M.~W., {Cook}, D.~A., {Doose}, L.~R., {Galuszka}, D.~M., {Hare},
  T.~M., {Howington-Kraus}, E., {Karkoschka}, E., {Kirk}, R.~L., {Lunine},
  J.~I., {McFarlane}, E.~A., {Redding}, B.~L., {Rizk}, B., {Rosiek}, M.~R.,
  {See}, C., {Smith}, P.~H., Nov. 2007. {Topography and geomorphology of the
  Huygens landing site on Titan}. \planss 55, 2015--2024.

\bibitem[{{Tomasko} et~al.(2005){Tomasko}, {Archinal}, {Becker}, {B{\'e}zard},
  {Bushroe}, {Combes}, {Cook}, {Coustenis}, {de Bergh}, {Dafoe}, {Doose},
  {Dout{\'e}}, {Eibl}, {Engel}, {Gliem}, {Grieger}, {Holso}, {Howington-Kraus},
  {Karkoschka}, {Keller}, {Kirk}, {Kramm}, {K{\"u}ppers}, {Lanagan},
  {Lellouch}, {Lemmon}, {Lunine}, {McFarlane}, {Moores}, {Prout}, {Rizk},
  {Rosiek}, {Rueffer}, {Schr{\"o}der}, {Schmitt}, {See}, {Smith}, {Soderblom},
  {Thomas}, and {West}}]{T05}
{Tomasko}, M.~G., {Archinal}, B., {Becker}, T., {B{\'e}zard}, B., {Bushroe},
  M., {Combes}, M., {Cook}, D., {Coustenis}, A., {de Bergh}, C., {Dafoe},
  L.~E., {Doose}, L., {Dout{\'e}}, S., {Eibl}, A., {Engel}, S., {Gliem}, F.,
  {Grieger}, B., {Holso}, K., {Howington-Kraus}, E., {Karkoschka}, E.,
  {Keller}, H.~U., {Kirk}, R., {Kramm}, R., {K{\"u}ppers}, M., {Lanagan}, P.,
  {Lellouch}, E., {Lemmon}, M., {Lunine}, J., {McFarlane}, E., {Moores}, J.,
  {Prout}, G.~M., {Rizk}, B., {Rosiek}, M., {Rueffer}, P., {Schr{\"o}der},
  S.~E., {Schmitt}, B., {See}, C., {Smith}, P., {Soderblom}, L., {Thomas}, N.,
  {West}, R., Dec. 2005. {Rain, winds and haze during the Huygens probe's
  descent to Titan's surface}. \nat 438, 765--778.

\bibitem[{{Tomasko} et~al.(2002){Tomasko}, {Buchhauser}, {Bushroe}, {Dafoe},
  {Doose}, {Eibl}, {Fellows}, {Farlane}, {Prout}, {Pringle}, {Rizk}, {See},
  {Smith}, and {Tsetsenekos}}]{T02}
{Tomasko}, M.~G., {Buchhauser}, D., {Bushroe}, M., {Dafoe}, L.~E., {Doose},
  L.~R., {Eibl}, A., {Fellows}, C., {Farlane}, E.~M., {Prout}, G.~M.,
  {Pringle}, M.~J., {Rizk}, B., {See}, C., {Smith}, P.~H., {Tsetsenekos}, K.,
  Jul. 2002. {The Descent Imager/Spectral Radiometer (DISR) Experiment on the
  Huygens Entry Probe of Titan}. \ssr 104, 469--551.

\bibitem[{{Tomasko} et~al.(2008){Tomasko}, {Doose}, {Engel}, {Dafoe}, {West},
  {Lemmon}, {Karkoschka}, and {See}}]{T08}
{Tomasko}, M.~G., {Doose}, L., {Engel}, S., {Dafoe}, L.~E., {West}, R.,
  {Lemmon}, M., {Karkoschka}, E., {See}, C., Apr. 2008. {A model of Titan's
  aerosols based on measurements made inside the atmosphere}. \planss 56,
  669--707.

\bibitem[{{Underwood}(1997)}]{U97}
{Underwood}, J.~C., 1997. {A system drop test of the Huygens Probe}. In: 14th
  Aerodynamic Decelerator Systems Technology Conference AIAA-1997-1429.

\bibitem[{{Witasse} et~al.(2008){Witasse}, {Huber}, {Zender}, {Lebreton},
  {Beebe}, {Heather}, {Matson}, {Zarnecki}, {Wheadon}, {Trautner}, {Tomasko},
  {Leon Stoppato}, {Simoes}, {See}, {Perez-Ayucar}, {Pennanech}, {Niemann},
  {McFarlane}, {Leese}, {Kazeminejad}, {Israel}, {Hathi}, {Hagermann},
  {Haberman}, {Fulchignoni}, {Ferri}, {Dutta-Roy}, {Doose}, {Demick-Montelara},
  {Colombatti}, {Brun}, {Bird}, {Atkinson}, and {Aboudan}}]{W08}
{Witasse}, O., {Huber}, L., {Zender}, J., {Lebreton}, J.-P., {Beebe}, R.,
  {Heather}, D., {Matson}, D.~L., {Zarnecki}, J., {Wheadon}, J., {Trautner},
  R., {Tomasko}, M., {Leon Stoppato}, P., {Simoes}, F., {See}, C.,
  {Perez-Ayucar}, M., {Pennanech}, C., {Niemann}, H., {McFarlane}, L., {Leese},
  M., {Kazeminejad}, B., {Israel}, G., {Hathi}, B., {Hagermann}, A.,
  {Haberman}, J., {Fulchignoni}, M., {Ferri}, F., {Dutta-Roy}, R., {Doose}, L.,
  {Demick-Montelara}, J., {Colombatti}, G., {Brun}, J.-F., {Bird}, M.,
  {Atkinson}, D., {Aboudan}, A., Apr. 2008. {The Huygens scientific data
  archive: Technical overview}. \planss 56, 770--777.

\bibitem[{{Zarnecki} et~al.(2002){Zarnecki}, {Leese}, {Garry}, {Ghafoor}, and
  {Hathi}}]{Z02}
{Zarnecki}, J.~C., {Leese}, M.~R., {Garry}, J.~R.~C., {Ghafoor}, N., {Hathi},
  B., Jul. 2002. {Huygens' Surface Science Package}. \ssr 104, 593--611.

\bibitem[{{Zarnecki} et~al.(2005){Zarnecki}, {Leese}, {Hathi}, {Ball},
  {Hagermann}, {Towner}, {Lorenz}, {McDonnell}, {Green}, {Patel}, {Ringrose},
  {Rosenberg}, {Atkinson}, {Paton}, {Banaszkiewicz}, {Clark}, {Ferri},
  {Fulchignoni}, {Ghafoor}, {Kargl}, {Svedhem}, {Delderfield}, {Grande},
  {Parker}, {Challenor}, and {Geake}}]{Z05}
{Zarnecki}, J.~C., {Leese}, M.~R., {Hathi}, B., {Ball}, A.~J., {Hagermann}, A.,
  {Towner}, M.~C., {Lorenz}, R.~D., {McDonnell}, J.~A.~M., {Green}, S.~F.,
  {Patel}, M.~R., {Ringrose}, T.~J., {Rosenberg}, P.~D., {Atkinson}, K.~R.,
  {Paton}, M.~D., {Banaszkiewicz}, M., {Clark}, B.~C., {Ferri}, F.,
  {Fulchignoni}, M., {Ghafoor}, N.~A.~L., {Kargl}, G., {Svedhem}, H.,
  {Delderfield}, J., {Grande}, M., {Parker}, D.~J., {Challenor}, P.~G.,
  {Geake}, J.~E., Dec. 2005. {A soft solid surface on Titan as revealed by the
  Huygens Surface Science Package}. \nat 438, 792--795.

\bibitem[{{Zebker} et~al.(2009){Zebker}, {Stiles}, {Hensley}, {Lorenz}, {Kirk},
  and {Lunine}}]{Z09}
{Zebker}, H.~A., {Stiles}, B., {Hensley}, S., {Lorenz}, R., {Kirk}, R.~L.,
  {Lunine}, J., May 2009. {Size and Shape of Saturn's Moon Titan}. Science 324,
  921--.

\end{thebibliography}

\newpage
\clearpage

\begin{table}
\centering
\caption{Instrument acronyms.}
\vspace{5mm}
\begin{tabular}{ll}
\hline
Acronym & Description \\
\hline
{\bf ACP} & Aerosol Collector and Pyrolyzer \\
{\bf AGC} & Automatic Gain Control \\
{\bf DISR} & Descent Imager / Spectral Radiometer \\
{\bf DLIS} & Downward Looking Infrared Spectrometer \\
{\bf DLV} & Downward Looking Violet photometer \\
{\bf DLVS} & Downward Looking Visual Spectrometer \\
{\bf DWE} & Doppler Wind Experiment \\
{\bf GCMS} & Gas Chromatograph / Mass Spectrometer \\
{\bf HASI} & Huygens Atmospheric Structure Instrument \\
{\bf MRI} & Medium Resolution Imager \\
{\bf RASU} & Radial Accelerometer Sensor Unit \\
{\bf SSL} & Surface Science Lamp \\
{\bf SSP} & Surface Science Package \\
{\bf ULIS} & Upward Looking Infrared Spectrometer \\
{\bf ULV} & Upward Looking Violet photometer \\
\hline
\end{tabular}
\label{tab:acronyms}
\end{table}

\newpage
\clearpage

\end{document}